\documentclass[aps,preprint,pre,groupedaddress,showpacs]{revtex4}
\usepackage{amsmath,amsbsy,graphicx}
\usepackage[section]{placeins}
\usepackage[usenames]{color}

\newcommand{\qh}{\hat{q}}
\newcommand{\hb}{\bar{h}}
\newcommand{\lr}[1]{\left.#1\right.}
\newcommand{\dhb}{\delta\bar{h}}
\newcommand{\du}{\delta u}

\begin{document}
\title{Undulation instability in a bilayer lipid membrane due to electric field
interaction with lipid dipoles}
\author{Richard J. Bingham}
\email{r.j.bingham01@leeds.ac.uk}
\affiliation{Polymers \& Complex Fluids Group, School of Physics \& Astronomy, University of Leeds, Leeds, LS2 9JT, United Kingdom}
\author{Stephen W. Smye}
\affiliation{Academic Division of Medical Physics, University of Leeds, Leeds, LS2 9JT, United Kingdom}
\author{Peter D. Olmsted}
\affiliation{Polymers \& Complex Fluids Group, School of Physics \& Astronomy, University of Leeds, Leeds, LS2 9JT, United Kingdom}
\begin{abstract}
Bilayer lipid membranes (BLMs) are an essential component of all biological systems, forming a functional barrier for cells and organelles from the surrounding environment.  The lipid molecules that form membranes contain both permanent and induced dipoles, and an electric field can induce the formation of pores when the transverse field is sufficiently strong (electroporation).  Here, a phenomenological free energy is constructed to model the response of a BLM to a transverse static electric field.  The model contains a continuum description of the membrane dipoles and a coupling between the headgroup dipoles and the membrane tilt.  The membrane is found to become unstable through buckling modes, which are weakly coupled to thickness fluctuations in the membrane.  The thickness fluctuations, along with the increase in interfacial area produced by membrane buckling, increase the probability of localised membrane breakdown, which may lead to pore formation.  The instability is found to depend strongly on the strength of the coupling between the dipolar headgroups and the membrane tilt as well as the degree of dipolar ordering in the membrane.             
\end{abstract}
\pacs{87.50.cj,87.16.ad,46.70.Hg}
\maketitle
\section{Introduction}
When amphiphilic lipid molecules are dissolved in solution, the molecules can self-assemble into a bilayer structure with the hydrophilic headgroups of the molecule shielding the hydrophobic hydrocarbon tails from the surrounding water.  Many biological processes and components that occur in the cell depend on the membrane: membrane-bound proteins, endo/exocytosis, lipid rafts and ion channels are just a few examples \cite{Albertsbook2002}. The many different species of lipid found in a cell membrane share the same general structure: a polar headgroup attached to a non-polar hydrocarbon tail region.  When a bilayer lipid membrane (BLM) forms, the non-polar tails make up the core of the membrane, with the dipolar headgroups forming the membrane surface \cite{SeifertAiP1997}. Both parts of the molecule react to electric fields.  When a strong electric field is applied transversely across a membrane, reversible electric breakdown can occur.  The breakdown is characterised by an increase in the measured conductivity due to the rapid increase in the transit of ions across the membrane \cite{StampfliAABC1958,NeumannJMB1972}.  This increase in permeability is attributed to the development of transbilayer pores \cite{Chizmadzhev1B&B1979}, which may close upon removal of the electric field, allowing the membrane to recover.  This phenomenon has been termed electroporation \cite{WeaverIEEE2003,ChenMBEC2006}.  

The theoretical work on electroporation and electrical breakdown can be viewed as belonging to two distinct branches:  one approach uses the Smoluchowski equation to describe the evolution of a distribution of pores with an assumed energy for pore formation \cite{WeaverB&B1986,WeaverB&B1991,KrassowskaBioJ1999,JoshiPRE2001,JoshiPRE2002,KrassowskaPRE2003,KrassowskaBioJ2004,KrassowskaBioJ2007,BicoutPRE2006}.  A density of pores are modelled drifting through radius space as function of time, generated by a source term including the effect of the field.  This method has been successful at predicting pore radii, lifetimes and densities, but does not model the mechanism of pore formation \cite{ChenMBEC2006}.  A different approach is required to understand how pores form and what membrane properties inform this process.  The simplest approach is to coarse-grain the BLM to a continuum membrane driven unstable by an electric field.  Early work by Crowley \cite{CrowleyBioJ1973} modelled the hydrocarbon core of the membrane as a dielectric slab with finite shear modulus and finite elastic compressibility, but estimates a critical transmembrane voltage an order of magnitude larger than experimental values \cite{ChenMBEC2006}.  The model of Lewis \cite{LewisIEEE2003} also models the membrane as a dielectric slab, but includes a Maxwell stress tensor which relates the dielectric constant to strain in the membrane, however also finds a critical transmembrane voltage larger than those experimentally reported \cite{WeaverIEEE2003}, similar to Crowley.  

These models neglect the fluid bilayer structure of the membrane, and thus neglect important mechanical properties such as a vanishing shear modulus and the bending rigidity.  The Helfrich-Canham Hamiltonian and its variants \cite{SeifertAiP1997}  are frequently used to model conformational changes to the membrane.  Sens and Isambert \cite{IsambertPRL2002} adapted these methods and considered the minimisation of the difference between the stressed and unstressed areas of a membrane in an electric field.  The authors imposed a force from the electric field on an undulating membrane and calculated the unstable undulatory wavelength and corresponding growth rate, although the model used neglects any thickness variation in the membrane.  The stochastic thermal undulations proposed as a mechanism for pore formation by Movileanu \emph{et al.} \cite{PopescuBMB2006} are hindered by a large energy barrier (91$k_{B}T$) and neglect the effect of the field on the membrane.  Membranes are only nm in thickness and the process of membrane breakdown under electric fields occurs over short time scales, which makes experimental study of pore formation difficult. 

Molecular dynamics (MD) studies of membranes have been used extensively to study the electrical behaviour of membranes \cite{TielemanBBA1997,TarekBioJ2005,TielemanBMCBiochem2004,GurtovenkoBioJ2007}.  These simulations provide molecular-level detail on a picosecond time scale.  However MD can only simulate a very small area of membrane for a short time.  The transbilayer pores opened during electroporation can last for up to ms \cite{WeaverIEEE2003} before closing, which MD simulations cannot capture.  Experimental studies range from measurement of the transmembrane current \cite{BierPRE2002,ChizmadzhevBioJ2001} to conductivity measurements using salt-filled vesicles \cite{NeumannEBJ1998}.  Recent developments in video microscopy and fluorescence have enabled the direct visualisation of giant unilamellar vesicles exposed to an electric field \cite{TekleBioJ2001,RiskeBioJ2005,DimovaSM2007}, in which pores can be directly observed.  

In this work, we develop a comprehensive, mesoscopic analytical approach, which includes a mechanical coupling between the orientation of the dipole on the surface and the membrane surface tilt.  This should destabilise the membrane as the headgroups seek to align with the field, rather than shield the hydrophobic core of the membrane from the surrounding fluid, which is their equilibrium position.  As the headgroups tilt, the membrane will tilt to try to restore the equilibrium position, which can introduce an instability in the membrane not noted previously in the literature.  We study this instability by performing a linear stability analysis of the free energy.  This perturbative approach will not capture the inherently discontinuous process of pore formation, but will predict the onset of instability in the membrane.  The instability occurs through deformational modes involving thickness fluctuations in the membrane, which increases the probability of localised breakdown and therefore of pore formation.  Applying a field to the membrane breaks the up/down symmetry of the membrane therefore it is important we include a description of the bilayer which allows each monolayer to be independently deformable. 

In Section~\ref{method} we construct the free energy including terms associated with mechanical deformation and introduce the description of the dipolar headgroups.  Section~\ref{results} presents the qualitative analysis of the model, Section~\ref{numcal} presents results from numerical calculations, and we conclude the paper and discuss possible future work in Section~\ref{conclusions}.                

\section{Model\label{method}}
\subsection{Geometry}
\begin{figure}[ht]
	\includegraphics[width=\linewidth]{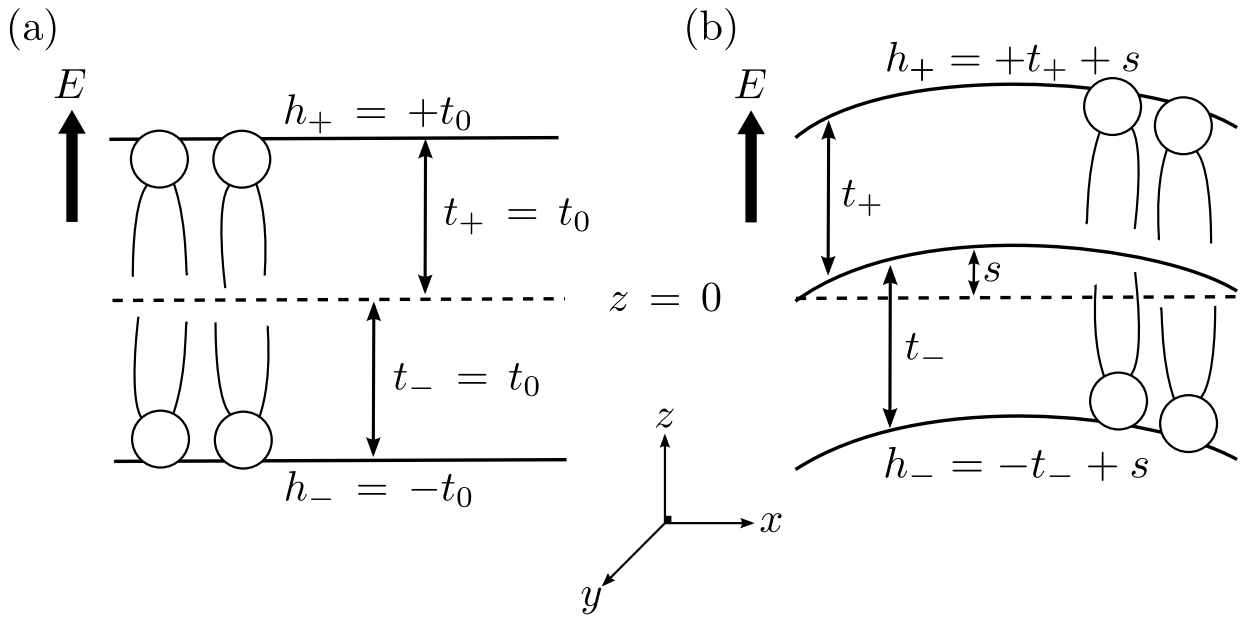}
	\caption{(a)An unperturbed bilayer, (b) the bilayer following a deformation.\label{pic2}}
\end{figure}

We consider a planar bilayer lipid membrane suspended horizontally in water with an electric field applied such that the field is perpendicular to the unperturbed membrane surface.  The membrane is modelled as a dielectric, fluid membrane at zero tension with a non-zero area stretching modulus.  The generalised three-dimensional (3D) free energy is given in Appendix~\ref{App:3d}, but to illustrate the basic principal and obtain analytically tractable solutions we assume a one-dimensional modulation in the $x$ direction (Figure~\ref{pic2}).  
\newline Here $h_{\pm}$ denote the positions of the upper (+) and lower (-) membrane surfaces, $t_{\pm}$ is the thickness of the upper and lower membrane leaflets, $t_{0}$ is the unperturbed monolayer thickness, and $s$ is the displacement of the dividing surface between the monolayers.  

\subsection{Conventional free energy}
We construct a phenomenological free energy per unit area.  The first contribution is the energy associated with mechanical deformation of the membrane;
\begin{equation}\label{fE}
f_{m}=\frac{\kappa_{b}}{2}\left(\left.h''_{+}\right.^{2}+\left.h''_{-}\right.^{2}\right)+\frac{\gamma}{2}\left(\left.h'_{+}\right.^{2}+\left.h'_{-}\right.^{2}\right)+\frac{\kappa_{A}}{2}\left[\left(\frac{t_{+}}{t_{0}}-1-t_{0}\,s''\right)^{2}+\left(\frac{t_{-}}{t_{0}}-1+t_{0}\,s''\right)^{2}\right]. 
\end{equation}
Here $\kappa_{A}$ is the area compressibility modulus, $\kappa_{b}$ is the bending rigidity, $\gamma$ is the surface tension and $t_{0}$ is the initial leaflet half-thickness.  The primes represent differentiation with respect to the $x$ direction.  These terms are equivalent to those used by Huang in \cite{HuangBioJ1986}, and have been adapted from the Helfrich-Canham Hamiltonian.  The surface tension in our model is not equivalent to a frame tension, which acts in the bilayer midplane.  Instead, it restricts variation in interfacial area of each leaflet separately and hence can describe peristaltic deformations.  The area compressibility term allows the two monolayers that form the bilayer to be independently deformable \cite{SeifertEPL1993}.  This has a strong effect on the relaxational dynamics of the membrane, but in the static case considered here these deformations will equilibrate effectively instantaneously, meaning we can minimise over $s$ at this stage without loss of generality.  [Note that the large bending modulus  ($\kappa_{b}\approx10\,k_{b}T$) precludes renormalisation of these elastic constants \cite{GoldsteinJPF1996}.] 

We also include the dielectric energy $f_{d}$: 
\begin{equation}\label{fD}
f_{d}=\frac{\epsilon_{0}}{2}\left[\epsilon_{m}E^{2}_{m}\left(h_{+}-h_{-}\right)+\epsilon_{w}E^{2}_{+}\left(L-h_{+}\right)+\epsilon_{w}E^{2}_{-}\left(h_{-}-L\right)\right].
\end{equation}
Here $\epsilon_{0}$ is the dielectric constant of the vacuum, $E_{m}$ is the field in the membrane, and $\epsilon_{m}$ and $\epsilon_{w}$ are the dielectric constants of the membrane and water respectively. $E_{\pm}$ is the field at the upper (+) and lower (-) membrane surface.  $L$ is the upper and lower limit of the system.  The form of Eq.~(\ref{fD}) implies the field will cause a uniform compressive force on the membrane.  This `electrostrictive' force has been found to have a quadratic dependence on the applied transmembrane voltage and has a small effect ($\sim1\%$ fractional thickness change) for the voltages used here \cite{EvansBioJ1975,RequenaBioJ1975,AndrewsSDFS1970}.            
 
\subsection{Dipolar Headgroups}
The dipolar headgroups of the lipids are defined by a three dimensional vector $\mathbf{p}$; 
\begin{eqnarray}\label{p:def}
\mathbf{p}&=&\mathbf{p}_{z}+\mathbf{m}\nonumber\\
&=&p\cos\left(\theta\right)\,\hat{e}_{z}+\mathbf{m},
\end{eqnarray}
where $p=\left|\mathbf{p}\right|$, $\theta$ is the angle the dipole makes with the $z$ direction, $\hat{e}_{z}$ is the unit vector in the $z$ direction and $\mathbf{m}$ is a vector representing the in-plane dipole moment.  The value used for $p$ is the effective magnitude of the headgroup dipole moment, calculated by Raudino and Mauzerall \cite{RaudinoBioJ1986}, which includes the screened charges and conformation of the headgroup.       

In an unperturbed bilayer lipid membrane, the headgroup of each phospholipid molecule lies at an average angle of $\theta_{0}$ to the membrane normal, hinged about the uppermost carbon atom \cite{BockmannBioJ2008}.  This natural tilt of the headgroup arises from a balance between the dipole-dipole interactions, the shape of the molecule and the need to shield the non-polar hydrocarbon chains from the water.  We perturb about the equilibrium position:  
\[
\theta_{\pm}=\theta_{0\pm}+\delta\theta_{\pm}
\]
where
\begin{eqnarray*}
\theta_{0+}&=&\theta_{0}\\
\theta_{0-}&=&\pi\pm\theta_{0}\qquad \left(A\,\,\text{and}\,\,B\right). 
\end{eqnarray*}

\begin{figure}[ht]
	\includegraphics[width=\linewidth]{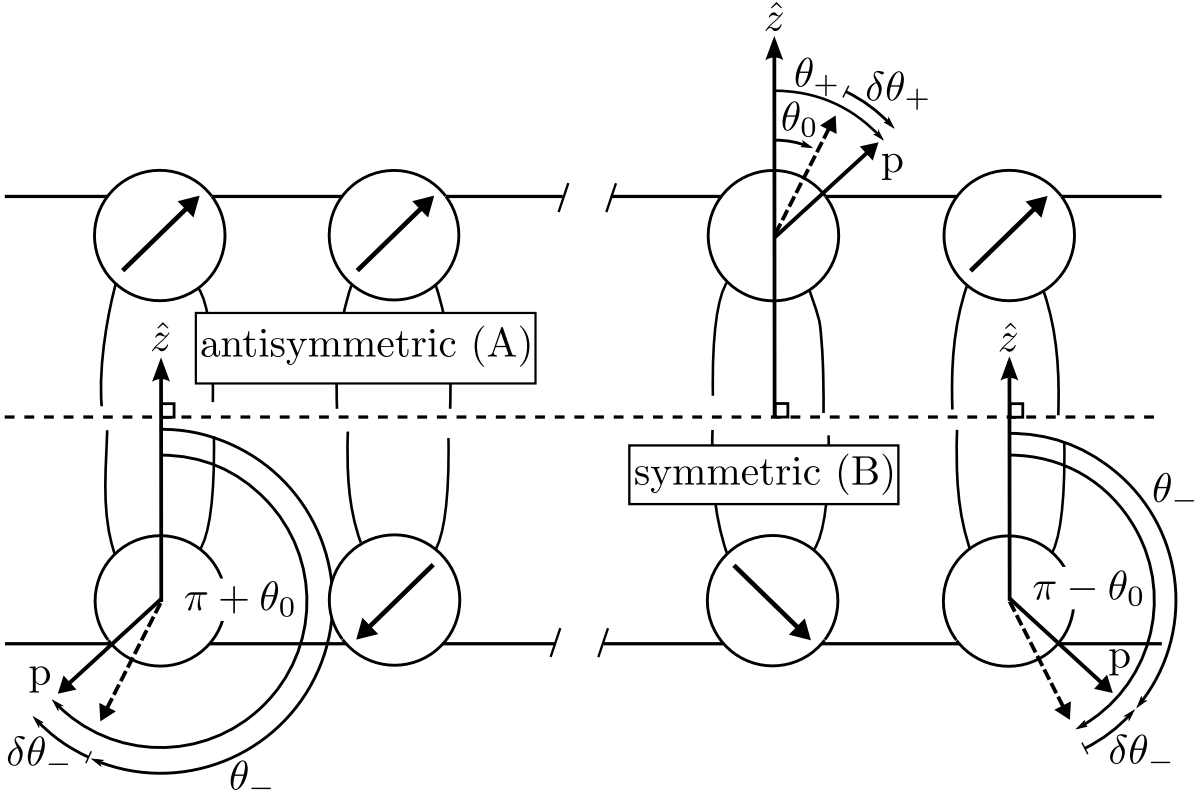}
	\caption{\label{pic3}Symmetric \& antisymmetric dipolar orientation between leaflets.}
\end{figure}

The dipolar orientations between the two leaflets are weakly coupled by the Coulomb interaction, which prefers antiparallel orientations.  In principle this degree of freedom allows for rich phase behaviour and dynamics.  Since the dipole orientation is coupled to the applied field, the relative orientation of the dipoles on the upper and lower membrane leaflets will affect the membrane behaviour under an electric field.  Here we consider only parallel and antiparallel orientations.  We refer to the case of the dipoles pointing in opposite directions as antisymmetric $\left(A\right)$, and the case where the dipoles point in the same direction as symmetric $\left(B\right)$, as shown in Figure~\ref{pic3}.      

Tilting the dipole relative to the membrane surface will cost energy, reflected in the following free energy per dipole;      
\begin{eqnarray}
\label{hydroE}
f_{p}&=&\frac{\kappa_{p}\left(\mathbf{m}_{+}\cdot\hat{e}_{x}\right)^{2}}{2}\left(\theta_{+}-\theta_{0+}+h'_{+}\right)^{2}+\frac{\kappa_{p}\left(\mathbf{m}_{-}\cdot\hat{e}_{x}\right)^{2}}{2}\left(\theta_{-}-\theta_{0-}+h'_{-}\right)^{2}\nonumber\\
\\\nonumber
&=&\frac{\kappa_{p}\left(\mathbf{m}_{+}\cdot\hat{e}_{x}\right)^{2}}{2}\left(\delta\theta_{+}+h'_{+}\right)^{2}+\frac{\kappa_{p}\left(\mathbf{m}_{-}\cdot\hat{e}_{x}\right)^{2}}{2}\left(\delta\theta_{-}+h'_{-}\right)^{2}  
\end{eqnarray}
Here $\kappa_{p}$ is the dipole-membrane coupling modulus and $\hat{e}_{x}$ is the unit vector in the $x$ direction.  This term is inspired in part by the work of Lubensky, Chen and MacKintosh in \cite{ChenPRE1995,ChenPRE1996,LubenskyPRL1993}, and explicitly penalizes change in dipole orientation relative to the membrane surface tilt.

So far $\mathbf{p}$ and $\mathbf{m}$ have referred to a single dipole.  To treat larger membrane areas, we extend these to continuum variables:  $\mathbf{p}$ becomes a dipole moment density $\widetilde{p}$ (dipole moment per unit area, with dimensions $\text{Cm}^{-1}$).  We coarse-grain $\mathbf{m}$ into $\mathbf{\overline{m}}\equiv\left\langle\mathbf{m}\right\rangle$, the average orientation of dipoles within a small area.  To distinguish between changes in orientation of the dipoles and changes in alignment within the coarse-grained area, we separate $\mathbf{\overline{m}}$ into two components; a unit vector $\widehat{m}=\overline{\mathbf{m}}/\left|\overline{\mathbf{m}}\right|$, representing the average orientation and an amplitude $\widetilde{m}=\left|\overline{\mathbf{m}}\right|$, which represents the degree with which dipoles within the given area point along $\widehat{m}$.  If all the dipoles within the area point along $\widehat{m}$, then $\widetilde{m}=1$.  If $\widetilde{m}=0$ then the dipoles within the area are completely disordered.  As we are considering a one-dimensional modulation in the $x$ direction we can set $\widehat{m}=\hat{e}_{x}$ and allow $\widetilde{m}$ to vary.  We write $\widetilde{m}$ as an equilibrium value $\widetilde{m}_{0}$, and the deviation $\delta \widetilde{m}_{\pm}$, from this equilibrium value.
\[
\widetilde{m}_{\pm}=\widetilde{m}_{0}+\delta \widetilde{m}_{\pm}.
\]
The equilibrium value $\widetilde{m}_{0}$ arises from a competition between the dipole-dipole interactions and their thermal fluctuations.  We allow $\delta \widetilde{m}_{\pm}$ to vary independently between the leaflets.   

Changes in dipole alignment are penalised by a susceptibility $\chi_{m}$, leading to the free energy density
\begin{equation}\label{mE}
f_{\chi}=\frac{\chi_{m}}{2}\,\delta \widetilde{m}^{2}_{\pm} 
\end{equation}
A three-dimensional construction of $f_{\chi}$ is given in Appendix~\ref{App:3d}.                
As in Andelman \emph{et al.} \cite{AndelmanJCP1986} we include a term coupling the dipole alignment with the surface curvature:  
\begin{equation}\label{cE}
f_{c}=-\frac{\gamma_{c}}{2}\left[\left(h''_{+}\right)\delta \widetilde{m}'_{+}+\left(h''_{-}\right)\delta \widetilde{m}'_{-}\right]
\end{equation}           
where $\gamma_{c}$ is the relevant modulus.

The final contribution to the free energy, $f_{E}$ is a free energy of two uncoupled dipoles in an electric field.  The free energy per unit area is  
\begin{eqnarray}\label{fieldE}
f_{E_{A,B}}&=&-\widetilde{p}\,E_{+}\,\cos\theta_{+}-\widetilde{p}\,E_{-}\,\cos\theta_{-}\nonumber\\
\nonumber\\
&=&\begin{cases}\widetilde{p}\,E \left[\frac{\cos\left(\theta_{0}\right)}{2}\left(\delta\theta_{+}^{2}-\delta\theta_{-}^{2}\right)+\sin\left(\theta_{0}\right)\left(\delta\theta_{+}-\delta\theta_{-}\right)\right] \quad\text{antisymmetric case $\left(A\right)$}\\
\\
\widetilde{p}\,E \left[\frac{\cos\left(\theta_{0}\right)}{2}\left(\delta\theta_{+}^{2}-\delta\theta_{-}^{2}\right)+\sin\left(\theta_{0}\right)\left(\delta\theta_{+}+\delta\theta_{-}\right)\right] \quad\text{symmetric case $\left(B\right)$,}\end{cases}
\end{eqnarray}
where we have assumed equal fields in the upper and lower water regions, $E_{+}=E_{-}=E$.  Here we have also assumed that the field acting on the dipoles is equivalent to the field at the membrane surface, not the field in the membrane core, which differs by two orders of magnitude.  The dipoles are in an aqueous environment which is very different from the hydrocarbon core of the membrane \cite{LewisIEEE2003}.  The relative dielectric constant of bulk water is $\approx80$, whereas for the membrane interior it is $\approx3$.  The field is assumed to only act between the two membrane surfaces hence the effect of ionic screening can be neglected.  Here $\theta_{\pm}$ refers to an average tilt angle within the region of coarse graining. 

Now that the free energy has been constructed, it can be subjected to a linear stability analysis.  This will predict the onset of a static instability in the membrane.  In an experimental system, the instability is complicated by the dynamics of the surrounding fluid (likely to contain many ions, particularly near the membrane surface) and by the dynamics of the bilayer itself, which behaves as a two-dimensional fluid.  To capture the dynamical behaviour of the instability predicted by our model, we would need to include both the hydrodynamic flows of the fluid and membrane \cite{BrochardJDP1975,SeifertEPL1993} and the movement of charges in the solution \cite{LacosteEPL2007,AjdariPRL1995}.  These are both non-trivial extensions due to the explicit definition of the dipoles in our model Eq.~(\ref{hydroE}-\ref{fieldE}) and hence beyond the scope of this paper.

\section{Qualitative Analysis\label{results}}  
\subsection{Fluctuation free energy}
\begin{figure}[ht]
		\includegraphics[]{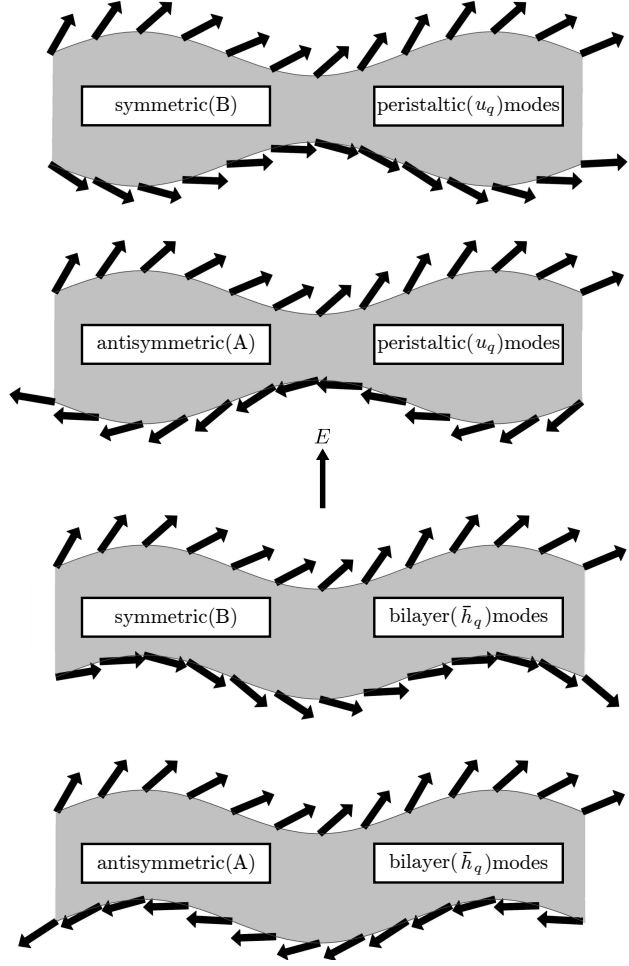}
		\caption{\label{dippic}The dipole orientations in an applied field.  The field applied to each membrane ($\phi=0.03$) is equivalent to a voltage drop of 0.07V across the membrane.  The parameters used are given in Table~\ref{tab2}.  The dipole orientations were generated using the equations for the minimums of $\Delta$ and $\Sigma$ found in Appendix~\ref{App:DS}.  The functional form of the bilayer deformation modes $u$ and $\bar{h}$ are imposed in order to display pure modes.}    
\end{figure}

After constructing the free energy we change variables to modes that characterize the bilayer as a whole:    
\begin{eqnarray*}
u&=&h_{+}-h_{-}\qquad\text{peristaltic mode}\\
\hb&=&h_{+}+h_{-}\qquad\text{bilayer mode}\\
\Delta&=&\delta\theta_{+}-\delta\theta_{-}\thinspace\quad\text{difference in dipole tilts}\\
\Sigma&=&\delta\theta_{+}+\delta\theta_{-}\thinspace\quad\text{mean dipole tilt.}\\
\end{eqnarray*}
From Equations (\ref{fE}-\ref{fieldE}), the free energy can then be expressed as;
\begin{eqnarray}
f&=&\frac{\kappa_{b}}{4}\left(\left.u''\right.^{2}+\left.\hb''\right.^{2}\right)+\frac{\gamma}{4}\left(\left.u'\right.^{2}+\left.\hb'\right.^{2}\right)+\frac{\chi_{m}}{2}\left(\delta \widetilde{m}_{+}^{2}+\delta \widetilde{m}_{-}^{2}\right)\nonumber\\
\nonumber\\
&+&\frac{\kappa_{A}}{2\,t_{0}^{2}}\left(\frac{u^{2}}{2}+2\,t_{0}^{2}-2\,t_{0}u\right)+\frac{\epsilon_{0}\epsilon_{w}\,E^{2}}{2}\left(\frac{\epsilon_{w}}{\epsilon_{m}}-1\right)u\nonumber\\
\nonumber\\
&+&\frac{\kappa_{p}\widetilde{m}^{2}_{0}}{4}\left(\Delta^{2}+\Sigma^{2}+\left.\hb'\right.^{2}+\left.u'\right.^{2}+2\,\hb'\Sigma+2\,u'\Delta\right)\nonumber\\
\\
&-&\frac{\gamma_{c}}{4}\left[\left(\hb''+u''\right)\delta \widetilde{m}'_{+}+\left(\hb''-u''\right)\delta \widetilde{m}'_{-}\right]\nonumber\\
\nonumber\\
&+&\begin{cases}+\widetilde{p}\,E\left(\frac{\cos\left(\theta_{0}\right)}{2}\Delta\,\Sigma+\sin\left(\theta_{0}\right)\Sigma\right)\quad\text{antisymmetric}\\
\nonumber\\+\widetilde{p}\,E\left(\frac{\cos\left(\theta_{0}\right)}{2}\Delta\,\Sigma+\sin\left(\theta_{0}\right)\Delta\right)\quad\text{symmetric.}\end{cases}
\end{eqnarray}
The system has six remaining degrees of freedom; $\delta\widetilde{m}_{\pm}$, $\Delta$, $\Sigma$, $u$ and $\bar{h}$.  We minimise $f$ over the dipolar tilts $\Delta$ \& $\Sigma$. The minimised values of these tilts are given in Appendix~\ref{App:DS}.  To quadratic order, the resulting free energy for the symmetric and antisymmetric case is identical.   Figure~\ref{dippic} shows the dipole configurations for the bilayer \& peristaltic modes.   
The variables $u$ and $\hb$ are expanded as small perturbations about the flat state;
\begin{equation}
u=u_{0}+\du \qquad\hb=\hb_{0}+\dhb
\end{equation}
where
\begin{equation}
\partial_{x}u_{0}=0 \qquad\partial_{x}\hb_{0}=0.
\end{equation}       
The equilibrium value of the peristaltic mode $u_{0}$ includes the electrostrictive thinning implied by Eq.~(\ref{fD}). 
 
The free energy will then consist of $f_{0}\left(u_{0},\bar{h}_{0}\right)$ and the perturbation $\delta f\left(\delta u, \delta\bar{h}\right)$.  The flat state $f_{0}\left(u_{0},\bar{h}_{0}\right)$ describes the membrane after the application of an electric field in the absence of undulations.  We expand the perturbation to second order in $\dhb$, $\du$ and $\delta m_{\pm}$. 
The stability of the system is determined by $\delta f$;
\begin{eqnarray}
\label{affe}
\delta f&=&\frac{\kappa_{b}}{4}\left(\lr{\dhb''}^{2}+\lr{\du''}^{2}\right)+\frac{\gamma}{4}\left(\lr{\du'}^{2}+\lr{\dhb'}^{2}\right)+\frac{\chi_{m}}{2}\left(\delta \widetilde{m}_{+}^{2}+\delta \widetilde{m}_{-}^{2}\right)\nonumber\\
\nonumber\\
&+&\frac{\kappa_{A}}{2\,t^{2}_{0}}\left(\frac{\lr{\du}^{2}}{2}\right)+\frac{\kappa_{p}\widetilde{m}^{2}_{0}}{4}\left(\lr{\dhb'}^{2}+\lr{\du'}^{2}\right)\nonumber\\
\\
&-&\frac{\left[\kappa_{p}^{3}\,\widetilde{m}^{6}_{0}\left(\lr{\du'}^{2}+\lr{\dhb'}^{2}\right)-2\,\kappa_{p}^{2}\,\widetilde{m}^{4}_{0}\,\widetilde{p}\,E\,\cos\left(\theta_{0}\right)\du'\,\dhb'\right]}{4\left(\kappa_{p}^{2}\,\widetilde{m}^{4}_{0}-\widetilde{p}^{2}\,E^{2}\,\cos^{2}\left(\theta_{0}\right)\right)}\nonumber\\
\nonumber\\
&-&\frac{\gamma_{c}}{4}\left[\left(\dhb''+\du''\right)\delta \widetilde{m}'_{+}+\left(\dhb''-\du''\right)\delta \widetilde{m}'_{-}\right].\nonumber
\end{eqnarray}
when $\delta f<0$ the system becomes unstable.

\subsection{Fourier Expansion}
Upon $\du$ and $\dhb$ decomposing into Fourier modes,
\begin{eqnarray}
&\du&=\sum_{q}u_{q}t_{0}e^{-i\qh x/t_{0}}\qquad\dhb=\sum_{q}\hb_{q}t_{0}e^{-i\qh x/t_{0}}\nonumber\\\\
&s&=\sum_{q}s_{q}t_{0}e^{-i\qh x/t_{0}}\qquad\delta \widetilde{m}_{\pm}=\sum_{q}\delta\widetilde{m}^{\pm}_{q}e^{-i\qh x/t_{0}}\nonumber
\end{eqnarray}
where $\qh$ is a normalised Fourier wavenumber given by $\qh=q\,t_{0}$,  The free energy fluctuation is given by 
\begin{equation}
\delta f=\frac{L^{2}}{\left(2\pi\right)^{2}}\left(\frac{\kappa_{b}}{2t^{4}_{0}}\right)\sum_{q}\delta f_{q} 
\end{equation}  
where  
\begin{eqnarray}
\label{finalfe}
\delta f_{q}&=&\frac{1}{2}\left[\qh^4+\left(\sigma_{s}-\sigma_{p}\,\frac{\phi^{2}}{\sigma_{p}^{2}-\phi^{2}}\right)\qh^{2}\right]\left(\lvert\hb_{q}\rvert^{2}+\lvert u_{q}\rvert^{2}\right)+\frac{\sigma_{A}}{2}\left(\lvert u_{q}\rvert^{2}\right)\nonumber\\
\nonumber\\
&+&\sigma_{p}^{2}\,\qh^{2}\frac{\phi}{\sigma_{p}^{2}-\phi^{2}}\left(\hb^{*}_{q}u_{q}+\hb_{q}u^{*}_{q}\right)+\sigma_{\chi}\left(\lvert\delta \widetilde{m}_{q}^{+}\rvert^{2}+\lvert\delta \widetilde{m}_{q}^{-}\rvert^{2}\right)\\
\nonumber\\
&+&\frac{i\,\sigma_{c}\,\qh^{3}}{2}\left[\left(\hb_{q}^{*}+u_{q}^{*}\right)\lr{\delta \widetilde{m}^{+}_{q}}+\left(\hb_{q}^{*}-u_{q}^{*}\right)\lr{\delta \widetilde{m}^{-}_{q}}\right].\nonumber
\end{eqnarray}
and
\begin{eqnarray*}
\sigma_{p}=\frac{\kappa_{p}\,\widetilde{m}^{2}_{0}\,t^{2}_{0}}{\kappa_{b}}\qquad\qquad&\sigma_{s}=\frac{\gamma\, t^{2}_{0}}{\kappa_{b}}&\qquad\qquad\sigma_{A}=\frac{\kappa_{A}\,t^{2}_{0}}{\kappa_{b}}\\
\sigma_{\chi}=\frac{\chi_{m}\,t^{2}_{0}}{\kappa_{b}}\qquad\qquad&\sigma_{c}=\frac{\gamma_{c}}{\kappa_{b}}&\qquad\qquad\phi=\frac{\widetilde{p}\,E\,\cos\left(\theta_{0}\right)t^{2}_{0}}{\kappa_{b}}\,\,.
\end{eqnarray*}

The dimensionless parameters reference all energies to the membrane bending energy $\kappa_{b}$.  The dimensionless potential $\phi$ compares the strength of the field on the dipoles with the membrane bending rigidity.  For $\phi\ll1$, we would therefore expect the system to be stable as the bending rigidity will be much stronger than the field on the dipoles.  We predict that the system will only become unstable for $\phi\approx O\left(1\right)$ where the strength of the field on the dipoles can overcome the bending rigidity.  The effect of the field through $\phi$ is regulated by $\sigma_{p}$; the two terms appear concurrently in Eq.~(\ref{finalfe}).  This is because $\sigma_{p}$ controls the way the membrane `feels' the field through the dipole-membrane coupling Eq.~(\ref{hydroE}).      

\subsection{Matrix Representation}
We rearrange Eq.~(\ref{finalfe}) in the quadratic form
\begin{equation}\label{quad.form}
\delta f_{q}=\frac{1}{2}\,\,\mathbf{v}^{T}\cdot\mathbf{M}_{q}\cdot\mathbf{v} \qquad\text{where}\,\,\mathbf{v}=\begin{pmatrix}u_{q},&\hb_{q},&\delta \widetilde{m}^{+}_{q},&\delta \widetilde{m}^{-}_{q}\end{pmatrix}
\end{equation}
where $\mathbf{v}$ is complex.  The form and components of $\mathbf{M}_{q}$ are given in Appendix~\ref{App:M}.  We calculate the eigenvalues $\lambda_{q}$  and eigenvectors $\hat{e}_{\lambda}$ of $\mathbf{M}_{q}$, which describe the eigenmodes of the system.  The membrane is unstable when any of the eigenvalues become negative. The associated eigenvectors $\hat{e}_{\lambda}$ determine how much the bilayer ($\bar{h}_{q}$), peristaltic ($u_{q}$) and dipole alignment ($\delta m^{\pm}_{q}$) modes are involved in each eigenmode.  If the eigenvector is dominated by one of these pure modes, we will refer to the eigenvalue as distinctly associated with the pure mode that dominates its eigenmode.  The matrix $\mathbf{M}_{q}$ can also be used to calculate the fluctuation spectrum of the modes;
\begin{equation}\label{fluc}
\left\langle\mathbf{v}_{i}\,\mathbf{v}_{j}\right\rangle=\alpha\left(\mathbf{M}_{q}^{-1}\right)_{ij}\,k_{B}T,
\end{equation}
where $\alpha=2\,\pi^{2}/L^{2}$ is a constant related to the Fourier expansion.

\subsection{Analytical Features}
We can gain some qualitative information about the membrane stability by analysing the fluctuation free energy, Eq.~(\ref{finalfe}).  The dimensionless parameters $\sigma_{A}$, $\sigma_{s}$, $\sigma_{p}$, $\sigma_{\chi}$ and $\sigma_{c}$ represent the effects of the area compressibility, surface tension, dipole-membrane coupling, dipole alignment and dipole alignment-membrane coupling on the free energy, scaled by the bending energy.  The typical values for these ratios, calculated using the parameters given in Table \ref{tab1}, are given in Table \ref{tab2}.  Since  $\sigma_{A}>1$, the membrane is more susceptible to bending than compression.  The ratio $\sigma_{p}$ controls how the membrane `feels' the applied electric field through the dipoles, as it contains both the dipole-membrane coupling modulus $\kappa_{p}$ and the average dipole alignment $\widetilde{m}_{0}$.  Since this ratio $\sigma_{p}$ is typically $<1$, this effect is dominated by membrane bending.  The ratio containing the surface tension, $\sigma_{s}$, which using the values of the parameters in Table~\ref{tab1} is $<1$.  This will cause the membrane to bend with surface gradients rather than surface curvature.   

The model should be stable ($\delta f_{q}\geq0$) for a flat membrane ($\hat{q}=0$).  For zero field ($\phi=0$) and in the limit $\qh\rightarrow 0$, the thickness variations (peristaltic modes $u_{q}$) are penalised by the area compressibility $\sigma_{A}$, while the bilayer modes vanish.  Since $u_{q}$ is stabilised against small changes in $\qh$, we expect the instability to progress via bilayer $\left(\bar{h}_{q}\right)$ modes.  If we consider only the bilayer modes, the instability occurs when the term proportional to $\hat{q}^{2}$, which can be considered to be an effective surface tension, changes sign:
\begin{equation}\label{phic}
\phi_{c}=\sigma_{p}\left(\frac{1}{1+\sigma_{p}/\sigma_{s}}\right)^{\frac{1}{2}}. 
\end{equation}
The critical potential $\phi_{c}$ is therefore approximately proportional to the ratio $\sigma_{p}$ \footnote{Although the linear dependence of $\phi_{c}$ on $\sigma_{p}$ is `softened' by the terms within the square root, the limiting behaviour of $\phi_{c}$ with respect to $\sigma_{p}$ is still preserved: $\phi_{c}\rightarrow\infty$ as $\sigma_{p}\rightarrow\infty$, and $\phi_{c}\rightarrow0$ as $\sigma_{p}\rightarrow0$.  The limiting case of $\phi_{c}=\sigma_{p}$ can only occur in the limit of infinite surface tension $\left(\sigma_{s}\rightarrow\infty\right)$ which means the apparent singularity in Eq.~(\ref{finalfe}) is always pre-empted by the membrane becoming unstable.  In the limit of zero surface tension, $\left(\sigma_{s}\rightarrow0\right)$ the critical potential $\phi_{c}\rightarrow0$ as there is nothing to resist the undulations introduced by the dipole-membrane coupling (Eq.~(\ref{cE})).}.  The ratio $\sigma_{p}$ contains two important independent parameters: $\kappa_{p}$, the strength of the dipole-membrane coupling and the average dipole alignment $\widetilde{m}_{0}$.  The membrane is stabilised upon increasing either $\kappa_{p}$ or $\widetilde{m}_{0}$.  Increasing $\kappa_{p}$ `stiffens' the dipoles against movement away from their equilibrium position, while increasing $\widetilde{m}_{0}$ increases the proportion of dipoles within a membrane patch that are aligned along the $x$ axis and therefore constrained by Eq.~(\ref{cE}).  As reducing $\widetilde{m}_{0}$ lowers $\phi_{c}$, we can infer that an instability is more likely to occur in a membrane region in which the dipoles are disordered (i.e. $\widetilde{m}_{0}$ is smaller).  The generalised 3D version of this term considers the orientation of the dipole in the complete plane, rather than just the $x$-direction, which could describe two dimensional modulations.  We expect this to be energetically more costly at the onset of instability \cite{ChenPRE1995}.
            
For the values used given in Table~\ref{tab2} $\left(\sigma_{s}=0.20,\sigma_{p}=0.23\right)$, $\phi_{c}\approx0.15$ which is equivalent to a voltage drop of $0.22\text{V}$ across the membrane.  This simple qualitative estimate gives values for the critical potential that are close to the experimentally reported range of (0.2-1V) \cite{WeaverIEEE2003}.  The inclusion of the coupling between the bilayer and peristaltic modes should raise the critical potential by transferring energy from the bilayer modes into the peristaltic modes.
 
\section{Numerical calculations\label{numcal}}
\subsection{Parameters}
All the parameters used to obtain these results are given in Table~\ref{tab1}.  The upper portion of the table contains the parameters that have been obtained from experimental studies, whereas those in the lower portion are not accessible by current experiments.  Those parameters that cannot be experimentally measured have either been estimated from physical reasoning ($\chi_{m}$, $\gamma_{c}$ and $\widetilde{m}_{0}$) or extrapolated from simulation results ($\kappa_{p}$).  Of these constants, we expect only $\kappa_{p}$ to have a significant effect on the stability of the membrane due it being present in Eq.~\ref{phic}.  The estimate of $\kappa_{p}$ comes from the distribution of dipole angles obtained by B\"{o}ckmann \emph{et al.} \cite{BockmannBioJ2008}.  The authors perform MD simulations of lipid bilayers and measure the angles formed between the dipole of the lipid headgroup and the surface normal.  This result has been produced in other studies, using different simulation methodologies \cite{BerkowitzBioJ2003,KleinJCP2002}.  We assume the width of this distribution is governed by one degree of freedom and then calculate the stiffness with which the dipole hinges around the equilibrium position, assuming it bends as a Hookean spring.
\begin{table*}[ht]
\caption{Parameters used for calculations}\label{tab1}
\begin{ruledtabular} 
\begin{tabular}{c c c c c} 
Symbol & Name & Value & Ref & Lipid \\[0.5ex]  \hline
$\kappa_{A}$ & Area compressibility & $0.14\,\text{Jm}^{-2}$ & \cite{Albertsbook2002} &POPC\\
$\kappa_{b}$ & Bending rigidity & $0.4\times10^{-19}\,\text{J}$ & \cite{Nelsonbook2004} &DMPC\\
$\gamma$ & Surface tension & $1.5\times10^{-3}\,\text{Jm}^{-2}$ & \cite{HuangBioJ1986} &GMO\\
$\theta_{0}$ & Equilibrium dipole orientation & $60^{\circ}$ & \cite{BockmannBioJ2008} &POPC\\
$\widetilde{p}$ & Dipole moment per unit area & $1.1\times10^{-9} \text{Cm}^{-1}$ & \cite{LewisIEEE2003} &POPC\\
$t_{0}$ & Monolayer thickness & $2.5\times10^{-9} \text{m}$ & \cite{Albertsbook2002} &POPC\\[0.5ex]\hline
$\kappa_{p}$ & Dipole-membrane coupling\footnote{The dipole-membrane coupling modulus, $\kappa_{p}$, was generated from the distribution of headgroup angles given in \cite{BockmannBioJ2008}.  The width of this distribution was assumed to be to be governed by one degree of freedom.  This estimate is then multiplied by the number of lipids per unit area, $n_{lip}$ ($\sim10^{19}$ $\text{lipids}/\text{m}^{2}$) \cite{Albertsbook2002}.}& $1.5\times10^{-2}\,\text{Jm}^{-2}$ & \cite{BockmannBioJ2008} &POPC\\
$\chi_{m}$ & Dipole alignment susceptibility\footnote{The dipole alignment susceptibility, $\chi_{m}$, was estimated as $\chi_{m}\sim k_{b}T\,n_{lip}$, due to the entropic cost of constraining the dipole alignment} & $3.45\times10^{-2}\,\text{Jm}^{-2}$ &-&-\\
$\gamma_{c}$ & Dipole alignment-membrane coupling\footnote{The functional form of $\sigma_{c}$ compares the dipole alignment-membrane coupling strength $\gamma_{c}$ with the bending rigidity $\kappa_{b}$.  As the effect that Eq.~(\ref{cE}) regulates is not seen experimentally, we can assume that $\gamma_{c}<\kappa_{b}$.  However the range of $\gamma_{c}$ is chosen in order to provide examples where $\gamma_{c}>\kappa_{b}$ as well as the physically expected range.} & $0.0-0.6\times10^{-19}\,\text{J}$ &-&-\\
$\widetilde{m}_{0} $ & Average degree of dipole alignment\footnote{The range of the dipolar alignment $\widetilde{m}_{0}$ used was judged to be a reasonable estimate of the degree of dipole alignment in a membrane.} & 0.3-0.4 &-&-\\
[1ex] 
\end{tabular}
\end{ruledtabular}
\end{table*}
\subsection{Eigenvalue stability}
\begin{figure}[ht]
	\includegraphics[width=\linewidth]{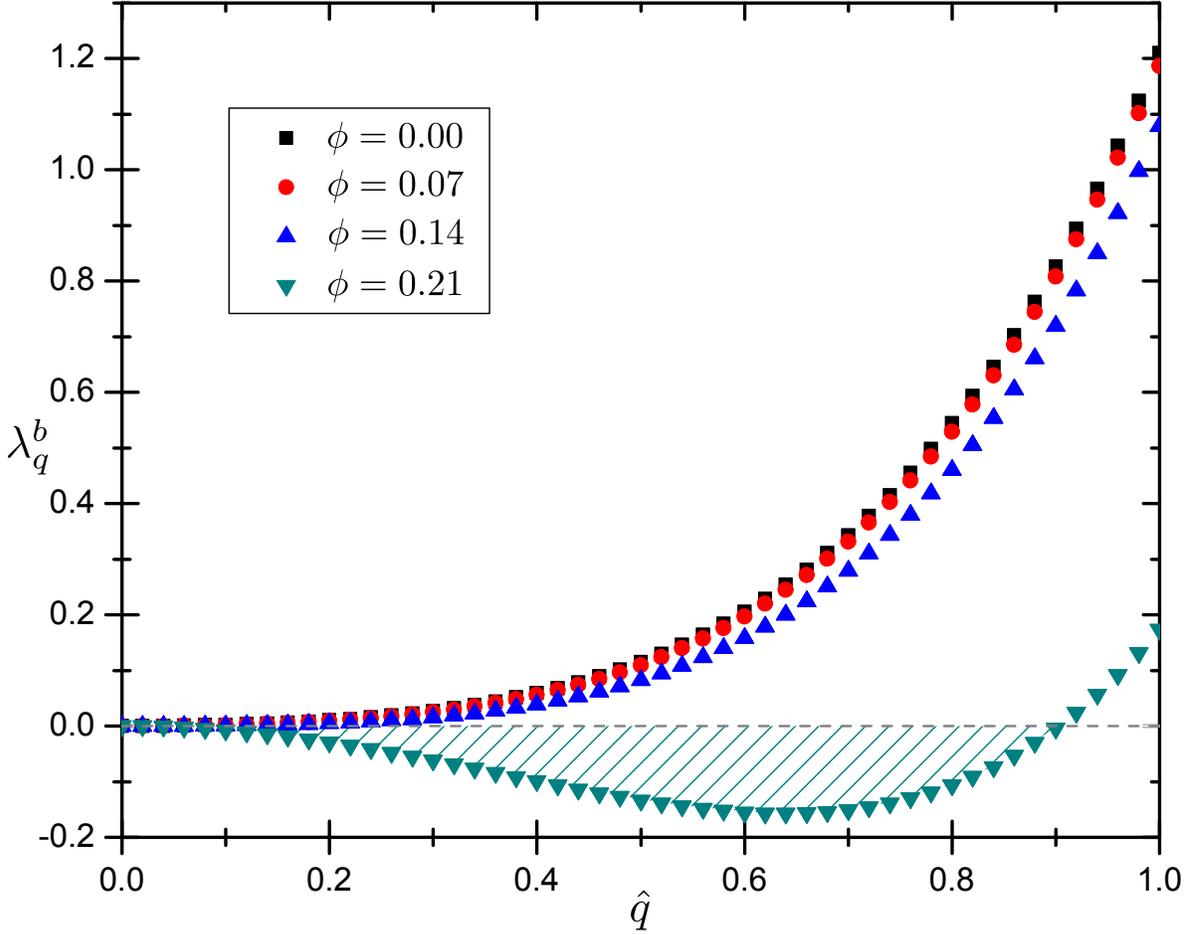}
  \caption{\label{graph1}The eigenvalue $\lambda_{q}^{b}$ associated with the bilayer undulations $\hb_{q}$ as a function of $\hat{q}$.  The shaded region is unstable.}
\end{figure}

Figure~\ref{graph1} shows the eigenvalue associated with the bilayer modes $\hb_{q}$.  As the rescaled potential $\phi$ is increased, the eigenvalue $\lambda^{b}_{q}$ decreases. For $\phi=0.21$ the eigenvalue has a significant negative portion indicating that $\phi$ has passed through the point at which the membrane first becomes unstable.  For this unstable value of $\phi$ the instability begins at $\hat{q}=0$ and reaches a lowest value at $\hat{q}=0.6$, which corresponds to peak to peak spacing of $26\text{nm}$.   The eigenvalue associated with the peristaltic undulations $u_{q}$ behaves identically to the eigenvalue for $\hb_{q}$, but has a $y$-intercept of $2 \sigma_{A}$, hence this branch will never become unstable for $\qh\in\left[0,1\right]$.  The membrane can therefore become unstable only through the bilayer modes of undulation, as suggested in the previous section.  While this is not an obvious route to transmembrane pore formation, bilayer modes have been observed to play an important role in MD simulations of electroporation \cite{TielemanBMCBiochem2004}, as well as occurring in the theoretical model of Sens and Isambert \cite{IsambertPRL2002}. 
                 
The locus of instability as a function of $\qh$ and $\phi$ is shown for two values of $\sigma_{p}$ in Figure~\ref{graph4}(a).  For $\sigma_{p}=0.23$ the membrane becomes unstable at the critical potential of $\phi_{c}=0.16$.  This corresponds to a voltage drop across the membrane of roughly $0.24\text{V}$,  which is in the range of values (0.2-1V) for the onset of electroporation seen in both experiments \cite{TekleBioJ2001} and simulations \cite{TarekBioJ2005}.  This is slightly higher than the qualitative estimate obtained in the previous section.  The instability in our model is likely to be relieved by a change of state of the membrane.  The formation of transmembrane pores can achieve this by allowing ions to permeate through the system, reducing the electric field across the membrane. 

Increasing the dipole-membrane coupling strength $\sigma_{p}$ to $0.34$ increases the critical potential to $\phi_{c}=0.22$ (0.34V).  This is again slightly larger than the value predicted by Eq.~(\ref{phic}) ($\phi_{c}=0.20$), but is also smaller than predicted by a linear increase of $\phi_{c}$ with $\sigma_{p}$.              
\subsection{Dipole alignment-membrane coupling}
\begin{figure}[ht]
	\includegraphics[width=\linewidth]{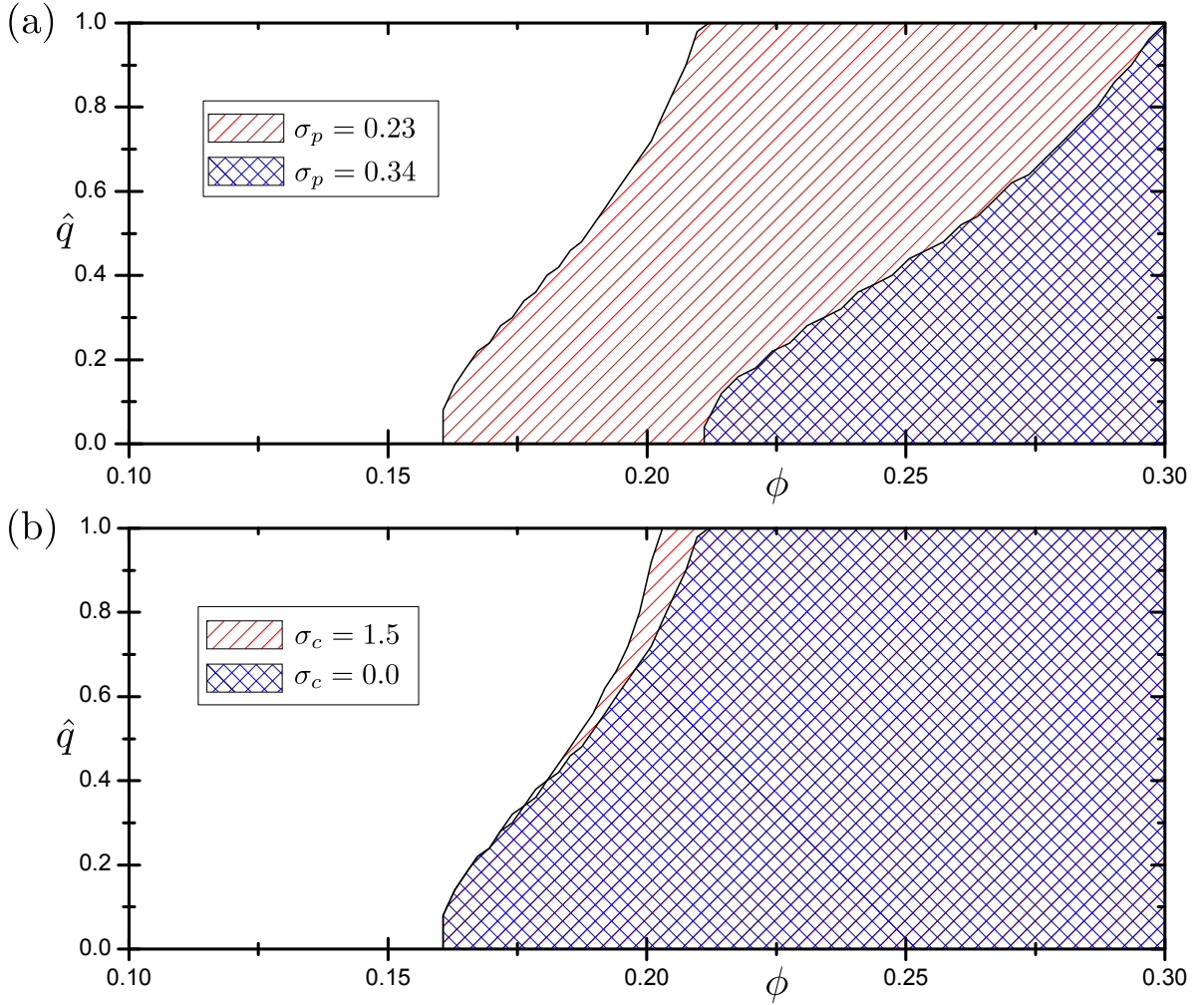}
	\caption{\label{graph4}The locus of instability as a function of $\hat{q}$ and $\phi$ for various values of (a) the dipole-membrane coupling $\sigma_{p}$ and (b) the dipole alignment-membrane coupling $\sigma_{c}$.  The shaded region is unstable.}
\end{figure}

The effect of the dipole alignment-membrane coupling $\sigma_{c}$ on the membrane stability is shown in Figure~\ref{graph4}(b). Increasing $\sigma_{c}$ does not affect the onset of the instability at long undulatory wavelengths ($\hat{q}=0$), where the cubic dependence on the wavelength ($\qh^{3}$) of $\sigma_{c}$ in the free energy is outweighed by the quadratic dependence ($\qh^{2}$) of the terms which cause the instability.  The variation of the dipole alignment-membrane coupling $\sigma_{c}$ does affect the shorter wavelength ($\qh\rightarrow1$) structure of the instability, where the cubic and quadratic terms become comparable.  This will affect the behaviour of the membrane if a field larger than the critical value is applied rapidly.  Overall $\gamma_{c}$ only weakly affects the stability of the membrane, over the range $\gamma_{c}=0-1.5\,\kappa_{b}$.  

\begin{figure}[ht]
	\includegraphics[width=0.7\linewidth]{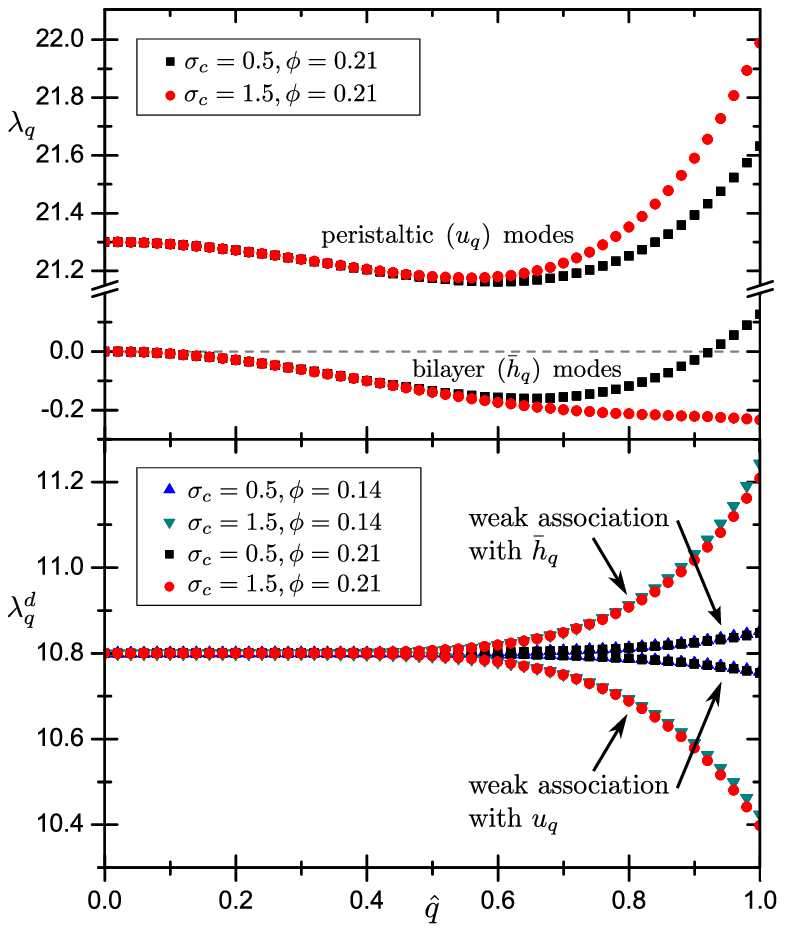}
  \caption{\label{graph6}The eigenvalues for both the peristaltic and bilayer modes (upper panel) and the dipole modes (lower panel) as functions of $\hat{q}$.  As $\sigma_{c}$ is increased at constant $\phi$ (=0.21), the bilayer and peristaltic modes react differently.  The lower branch of the bilayer modes becomes stable for $\qh>1$.  The dipolar modes show little difference for field strengths above or below the critical field strength $\left(\phi_{c}=0.16\right)$, but respond strongly to changes in the dipole alignment-membrane coupling $\sigma_{p}$.  As $\hat{q}$ is increased, the dipolar modes deviate from their initial value, the direction determined by their weak association with either the bilayer modes $\bar{h}_{q}$ or the peristaltic modes $u_{q}$.}
\end{figure}

Figure~\ref{graph4}(b) shows the effect of the dipole alignment-membrane coupling $\sigma_{c}$ on the stability of the membrane and thus only the effect of $\sigma_{c}$ on the bilayer modes.  The variation of $\sigma_{c}$ affects the peristaltic $\left(u_{q}\right)$ modes differently, as shown in Figure~\ref{graph6}.  The peristaltic modes are stabilised at higher $\hat{q}$ by an increase in the dipole alignment-membrane coupling $\sigma_{c}$, whereas the bilayer modes are destabilised.  

The dipole alignment-membrane coupling energy (Eq.~\ref{cE}) couples the membrane deformation modes ($\bar{h}_{q}$ and $u_{q}$) with the dipole alignment modes $\delta m^{\pm}_{q}$. The eigenvalues $\lambda_{q}^{d}$, whose eigenvectors are dominated by the dipole alignment modes $\delta m^{\pm}_{q}$ are shown in the lower section of Figure~\ref{graph6}.  The $\hat{q}=0$ limit of these eigenvalues is governed by and proportional to $\sigma_{\chi}$, and the eigenvalues only  deviate from this value at larger $\qh$.  Significant change in the eigenvalues can only be seen for values of $\sigma_{c}>1$.  These large values of $\sigma_{c}$ are unlikely to be physically realisable as they require $\gamma_{c}\approx O\left(\kappa_{b}\right)$.  The magnitude of $\gamma_{c}$ cannot be measured directly by current experiment.  However as the degree of dipole alignment is observed to have little correlation with membrane bending \cite{KotulskaBioelectro2007,KleinJCP2002} it is fair to assume $\gamma_{c}\ll\kappa_{b}$.

For the $\delta m_{+}$- and $\delta m_{-}$-dominated eigenvectors, the corresponding eigenvalues $\lambda_{q}^{d}$ are degenerate at $\hat{q}=0$ and then separate as $\hat{q}$ increases.  Increasing the strength of the dipole alignment-membrane coupling $\sigma_{c}$ increases the amount by which these eigenvalues deviate.  The corresponding eigenvectors are not associated with $\delta m_{+}$ and $\delta m_{-}$ independently, but rather with both modes equally; however the more stable eigenvalue has a slight contribution from the bilayer mode $\bar{h}_{q}$, whereas the less stable eigenvalue has a slight contribution from the peristaltic modes $u_{q}$.  The difference between these modes explains why the eigenvalues associated with the bilayer and peristaltic modes react differently to increases in $\sigma_{c}$

\begin{table}[ht]
\caption{Dimensionless ratios according to measured data from Table \ref{tab1}}\label{tab2}
\begin{tabular}{ccc}\hline\hline
Symbol & Formula & Value \\ [0.5ex]  \hline
$\sigma_{A}$ & $\frac{\kappa_{A}\,t^{2}_{0}}{\kappa_{b}}$ & $21.3$  \\\\
$\sigma_{s}$ & $\frac{\gamma\, t^{2}_{0}}{\kappa_{b}}$ & $0.2$  \\\\
$\sigma_{p}$ & $\frac{\kappa_{p}\,\widetilde{m}^{2}_{0}\,t^{2}_{0}}{\kappa_{b}}$ & $0.23-0.35$  \\\\
$\sigma_{\chi}$ & $\frac{\chi_{m}\,t^{2}_{0}}{\kappa_{b}}$ & $5.4$  \\\\
$\sigma_{c}$ & $\frac{\gamma_{c}}{\kappa_{b}}$ & $0-1.5$  \\\\
$\phi$ & $\frac{\widetilde{p}\,E\,\cos\left(\theta_{0}\right)t^{2}_{0}}{\kappa_{b}}$ & $0-0.35$ \\ 
[1ex]\hline
\end{tabular}
\end{table}
\subsection{Fluctuation spectrum}
\begin{figure}[ht]
	\includegraphics[width=\linewidth]{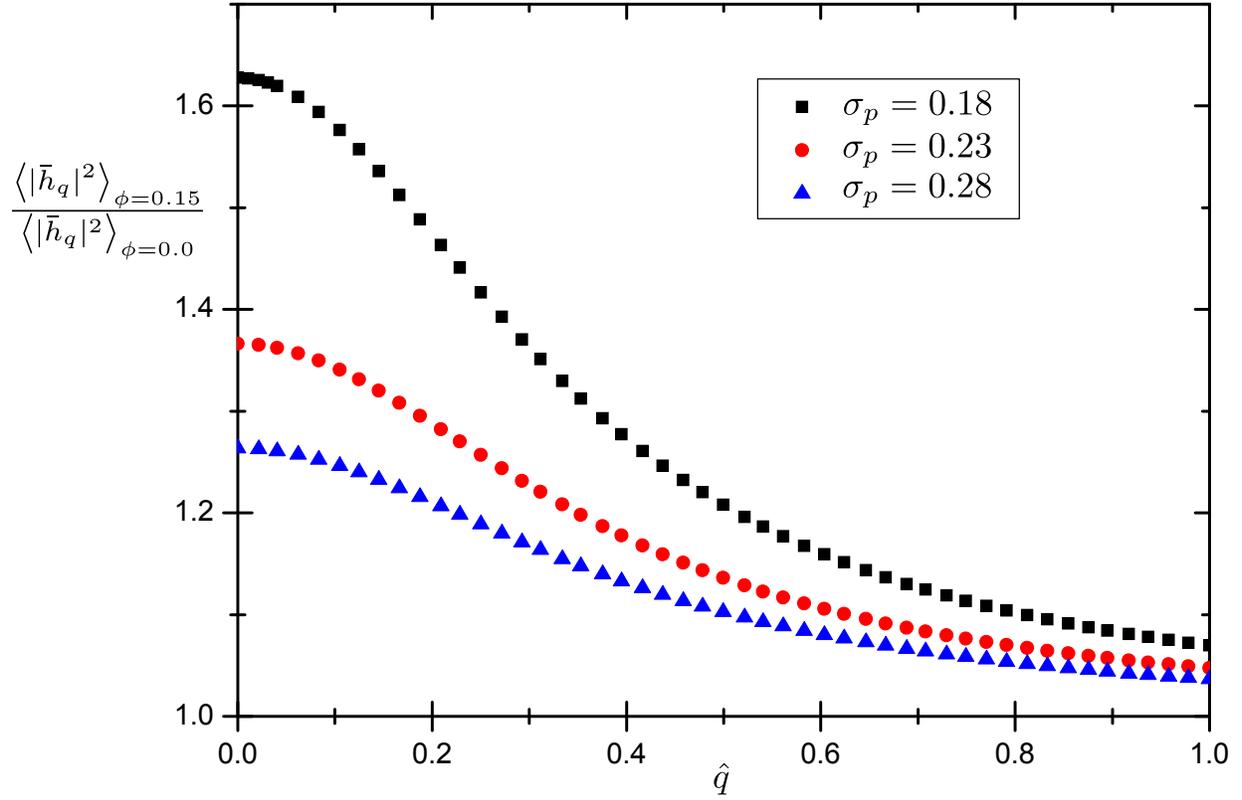}
  \caption{\label{graph5}The ratio between the fluctuations of the bilayer $\left(\bar{h}_{q}\right)$ modes at $\phi=0.15$ to $\phi=0$ as a function of $\hat{q}$.} 
\end{figure}

Whereas the variation of $\sigma_{c}$ has a small effect on the overall stability of the membrane, the interaction of $\sigma_{p}$ with the fluctuation modes provides a test of the model.  The fluctuation spectrum of the modes can be calculated using Eq.~(\ref{fluc}).  As expected, the fluctuations of the bilayer modes are much larger in magnitude than the peristaltic modes.  This is because the peristaltic modes are dominated by the strong stretching modulus, $\sigma_{A}$.  For small $\sigma_{A}$ the fluctuations of the peristaltic modes grow to match the fluctuations of the bilayer modes.  For zero applied field $\left\langle\lvert\hb_{q}\rvert^{2}\right\rangle\sim1/\left(\hat{q}^{4}+\sigma_{s}\hat{q}^{2}\right)$, as expected for a flat membrane \cite{SeifertAiP1997}.  The ratio of fluctuations of the bilayer modes at $\phi=0.15$ to $\phi=0.0$, $\left\langle\lvert\hb_{q}\rvert^{2}\right\rangle_{\phi=0.15}/\left\langle\lvert\hb_{q}\rvert^{2}\right\rangle_{\phi=0.0}$ is displayed in Figure~\ref{graph5}, for various values of the dipole-membrane coupling strength $\sigma_{p}$, showing that the application of a field increases the magnitude of the fluctuations.  Conversely, the dipole-membrane coupling $\sigma_{p}$ `stiffens' the membrane, and reduces the fluctuations as reflected in Figure~\ref{graph5}.
\begin{figure}[ht]
	\includegraphics[width=\linewidth]{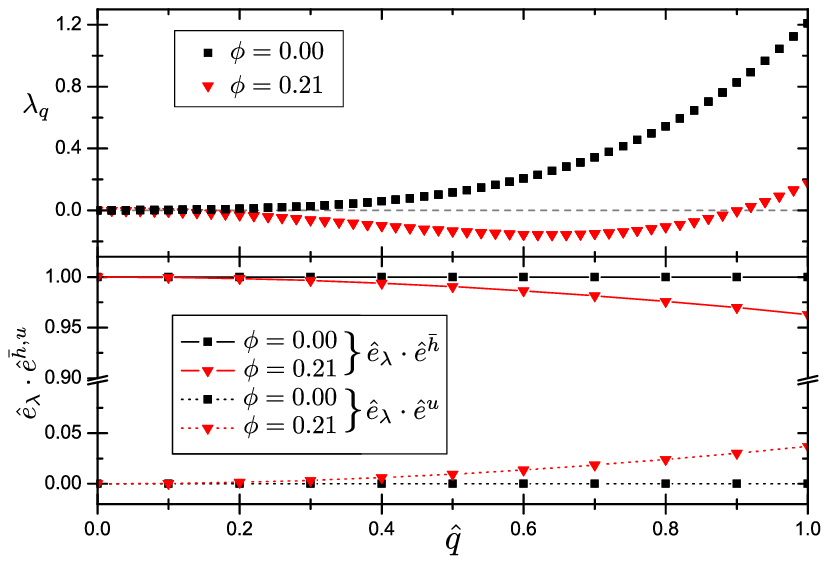}
  \caption{\label{evgraph}The unstable eigenvalue and corresponding eigenvector $\hat{e}_{\lambda}$ as a function of $\hat{q}$.  $\hat{e}_{\lambda}\cdot\hat{e}^{\bar{h}}$ and $\hat{e}_{\lambda}\cdot\hat{e}^{u}$ are the contributions to $\hat{e}_{\lambda}$ of the bilayer and peristaltic modes respectively.}
\end{figure}
\subsection{Eigenvector composition}
Figure~\ref{evgraph} shows the unstable eigenvalue and the associated normalised eigenvector.  For zero applied field, $\hat{e}_{\lambda}\cdot\hat{e}^{\bar{h}}=1$ and $\hat{e}_{\lambda}\cdot\hat{e}^{u}=0$, where $\hat{e}^{\bar{h}}$ and $\hat{e}^{u}$ are the unit vectors representing the pure modes $\bar{h}_{q}$ and $u_{q}$ respectively, hence the eigenvalue is associated only with the bilayer modes.  For $\phi>\phi_{c}$ the amplitudes of the eigenvector components vary with $\hat{q}$ and the contribution from the peristaltic mode $\hat{e}^{\lambda}\cdot\hat{e}^{u}_{q}$ increases due to the coupling term present in Eq.~(\ref{finalfe}).  This term has a similar field dependence to the effective surface tension term present in Eq.~(\ref{finalfe}) which induces the instability, so the eigenvector mixing increases dramatically for $\phi>\phi_{c}$.  The change in the eigenvector components is small compared with the initial composition, therefore we can consider the eigenvalue as distinctly associated the bilayer modes.  This behaviour is mirrored in the eigenvalues of the pure peristaltic modes, with the bilayer mode contribution ($\hat{e}_{\lambda}\cdot\hat{e}^{\bar{h}}$) increasing slightly for $\phi>\phi_{c}$ and increasing $\hat{q}$. 

\section{Discussion \& Summary\label{conclusions}}
We have constructed a model of a planar membrane in an electric field, which contains an explicit coupling between the orientation of the dipolar lipid headgroups and the membrane shape, thus coupling the application of the field to the membrane shape in a way not seen previously in the literature.  The phenomenological model contains only harmonic terms which are subjected to a linear stability analysis.  This model becomes unstable as the applied field is increased, with a critical potential that matches those seen in experiment and simulation \cite{ChenMBEC2006}.  A simple formula (Eq.~(\ref{phic})) has been found that gives a reasonable estimate of the critical potential related to a minimal number of model parameters, which is useful as decreasing the number of parameters used decreases possible sources of error.  The instability depends strongly on $\widetilde{m}_{0}$, the average alignment in a membrane patch, with the instability occurring for smaller fields for disordered membranes of smaller $\widetilde{m}_{0}$.  As dipole alignment will vary dynamically in a physical system, the membrane is more likely to become unstable in disordered patches.  This means the model captures some of the stochastic nature of membrane breakdown and pore formation.  The instability also depends strongly on $\kappa_{p}$, the strength of the coupling between the dipolar headgroups and the membrane core.  This is likely dependent on the combination of lipids in the bilayer.  Since variations in $\widetilde{m}_{0}$ or $\kappa_{p}$ have a significant effect on the critical potential $\phi_{c}$, these would be good parameters with which to test the model. 

Because the process of membrane breakdown requires a rupture to form in the membrane, it cannot be fully modelled by any continuum theory.  Despite this, the instability studied in this work can be linked with the formation of defects within the membrane and therefore the formation of transmembrane pores.  From Figure~\ref{evgraph}, the unstable eigenvector shows that the instability is dominated by the bilayer modes but approximately $2.5\%$ of the instability involves the peristaltic modes.  This induces a periodic thinning which destabilises the membrane.  Evans, Waugh and Melnick \cite{EvansBioJ1976} found using micropipette aspiration that a membrane can only support thickness changes of $\sim4\%$ before rupture.  To induce a fractional thickness change of this magnitude using the peristaltic undulations produced by the instability requires the bilayer modes to have an amplitude of $6\text{nm}$.  This is above the size that would be produced spontaneously by thermal fluctuations, but after the application of an electric field, the bilayer modes become unstable and this amplitude could be achieved.  A membrane defect is then more likely to form at the troughs of the peristaltic undulations, where the membrane is thinnest.  This defect could then go on to form a pore, or rupture the entire bilayer.  The most unstable undulation wavelength is $26\text{nm}$ (Figure~\ref{graph1}), comparable with the average pore-pore separation reported in \cite{FreemanBioJ1994}, consistent with this hypothesis.
       
The parameters $\kappa_{p}$ and $\gamma_{c}$, both unique to this model, provide opportunities to make predictions and test this model.  An obvious extension to the model would be to allow for the full rotation of the dipole distribution, which could lead to non-trivial pattern formation \cite{ChenPRE1995,ChenPRE1996,LubenskyPRL1993}.  Our calculations only calculate the static instability.   To fully model the dynamical behaviour of the instability predicted in this work, we would need to include both the hydrodynamic flows of the fluid and membrane \cite{BrochardJDP1975,SeifertEPL1993} and the movement of charges in the solution \cite{LacosteEPL2007,AjdariPRL1995}.  Coupling hydrodynamic flows to the movement of the membrane would be expected to push the instability to smaller wavelengths (larger $\hat{q}$) \cite{HohenbergRoMP1977}.                                                   

\begin{acknowledgments}
The authors would like to thank the EPSRC and the White Rose Doctoral Training Centre for funding.  R.B. would also like to thank Lisa Hawksworth and Jack Leighton for illuminating discussions.  
\end{acknowledgments}

\appendix
\section{3D free energy\label{App:3d}}
The general 3D form of the free energy of membrane deformation;
\begin{eqnarray}
f_{m}&=&\frac{\kappa_{b}}{2}\left(\left(\nabla^{2}h_{+}\right)^{2}+\left(\nabla^{2}h_{-}\right)^{2}\right)+\frac{\gamma}{2}\left(\left(\nabla h_{+}\right)^{2}+\left(\nabla h_{-}\right)^{2}\right)\nonumber\\
\\
&+&\frac{\kappa_{A}}{2}\left(\left(\frac{t_{+}}{t_{0}}-1-t_{0}\nabla^{2}s\right)^{2}+\left(\frac{t_{-}}{t_{0}}-1+t_{0}\nabla^{2}s\right)^{2}\right).\nonumber
\end{eqnarray}
For the free energy associated with the dipole surface coupling $f_{p}$ 
\begin{eqnarray}
f_{p}&=&\frac{\kappa_{p}}{2}\left\{\left[\left(\left(\mathbf{\hat{p}^*}_{+}-\mathbf{\hat{p}}_{+}\right)\cdot\hat{n}\right)\left|\mathbf{\hat{p}}_{+}\cdot\hat{\nabla h_{+}}\right|-\left(\mathbf{\hat{p}}_{+}\cdot\left(\nabla h_{+} - \nabla h_{0+}\right)\right)\left|\mathbf{\hat{p}}_{\bot+}\right|\right]^{2}\right.\nonumber\\
\\
&+&\left.\left[\left(\left(\mathbf{\hat{p}^*}_{-}-\mathbf{\hat{p}}_{-}\right)\cdot\hat{n}\right)\left|\mathbf{\hat{p}}_{-}\cdot\hat{\nabla h_{-}}\right|-\left(\mathbf{\hat{p}}_{-}\cdot\left(\nabla h_{-} - \nabla h_{0-}\right)\right)\left|\mathbf{\hat{p}}_{\bot-}\right|\right]^{2}\right\}\nonumber
\end{eqnarray}
where
\[
\mathbf{p}_{\bot}=\mathbf{\hat{p}}-\left(\mathbf{\hat{p}}\cdot\hat{n}\right)\hat{n}
\]
and the hatted variables are normalised.  $h_{0\pm}$ is the initial surface gradient and $\mathbf{p^*}$ is the perturbed dipole vector.  

$f_{c}$ is the free energy of the dipole alignment-membrane coupling ; 
\begin{equation}
f_{c}=\frac{\gamma_{c}}{2}\left(\left(\nabla^{2}h_{+}\,\nabla\cdot\mathbf{\hat{p}}_{+}\right)+\left(\nabla^{2}h_{-}\,\nabla\cdot\mathbf{\hat{p}}_{-}\right)\right).
\end{equation}
$f_{\chi}$ is the energy punishing dipole alignment;
\begin{equation}
f_{\chi}=\frac{\chi_{m}}{2}\left[\left(\mathbf{\hat{p}^*}_{\bot+}-\mathbf{\hat{p}}_{\bot+}\right)^{2}+\left(\mathbf{\hat{p}^*}_{\bot-}-\mathbf{\hat{p}}_{\bot-}\right)^{2}\right]
\end{equation}
$f_{d}$ has the same functional form but is integrated over three directions instead of two.   
\section{Matrix representation\label{App:M}}
The matrix constructed for Eq.~(\ref{quad.form}) is given by; 
\begin{equation}\label{M}
\mathbf{M}_{q}=\begin{Bmatrix}M_{11}&0&M_{13}&0&0&M_{16}&0&-M_{16}\\0&M_{11}&0&M_{13}&-M_{16}&0&M_{16}&0\\M_{13}&0&M_{33}&0&0&M_{16}&0&M_{16}\\0&M_{13}&0&M_{33}&-M_{16}&0&-M_{16}&0\\0&-M_{16}&0&-M_{16}&M_{55}&0&0&0\\M_{16}&0&M_{16}&0&0&M_{55}&0&0\\0&M_{16}&0&-M_{16}&0&0&M_{55}&0\\-M_{16}&0&M_{16}&0&0&0&0&M_{55}\end{Bmatrix}.  
\end{equation}
where
\begin{eqnarray*}
M_{11}&=&\qh^4+\left(\sigma_{s}-\sigma_{p}\frac{\phi^{2}}{\sigma_{p}^{2}-\phi^{2}}\right)\qh^{2}+\sigma_{A}\qquad M_{16}=\sigma_{c}\qh^{3}\\\\
M_{13}&=&\sigma_{p}^{2}\,\qh^{2}\frac{\phi}{\sigma_{p}^{2}-\phi^{2}}\thinspace\qquad\qquad\qquad\qquad\qquad M_{55}=2\sigma_{\chi}\\\\
M_{33}&=&\qh^4+\left(\sigma_{s}-\sigma_{p}\frac{\phi^{2}}{\sigma_{p}^{2}-\phi^{2}}\right)\qh^{2}.
\end{eqnarray*}
The vector $\mathbf{v}_{q}$ multiplying the matrix $\mathbf{M}_{q}$ consists of the real and imaginary parts of the modes $\hb_{q}$, $u_{q}$ and $\delta\widetilde{m}^{\pm}_{q}$;
\begin{equation}
\mathbf{v}=\begin{pmatrix}\left(u_{q}\right)_{r},&\left(u_{q}\right)_{i},&\left(\hb_{q}\right)_{r},&\left(\hb_{q}\right)_{i},&\left(\delta \widetilde{m}^{+}_{q}\right)_{r},&\left(\delta \widetilde{m}^{+}_{q}\right)_{i},&\left(\delta \widetilde{m}^{-}_{q}\right)_{r},&\left(\delta \widetilde{m}^{-}_{q}\right)_{i}\end{pmatrix}.
\end{equation}
\section{The minimised values $\Sigma_{min}$, $\Delta_{min}$ and $s_{q\,min}$ \label{App:DS}}
The minimised values of $\Sigma$, $\Delta$ and $s_{q}$ are given by;
\begin{equation}
\begin{pmatrix}
\Sigma\\
\Delta
\end{pmatrix}_{A}=\frac{1}{\sigma_{p}^{2}-\phi^{2}}\begin{pmatrix}\sigma_{p}\phi u'+2\,\phi^{2}\tan\left(\theta_{0}\right)-\sigma_{p}^{2}\bar{h}'\\\sigma_{p}\phi\bar{h}'-2\,\sigma_{p}\phi\tan\left(\theta_{0}\right)-\sigma_{p}^{2}u'\end{pmatrix}
\end{equation}
or
\begin{equation}
\begin{pmatrix}
\Sigma\\
\Delta
\end{pmatrix}_{B}=\frac{1}{\sigma_{p}^{2}-\phi^{2}}\begin{pmatrix}\sigma_{p}\phi u'-2\,\sigma_{p}\phi\tan\left(\theta_{0}\right)-\sigma_{p}^{2}\bar{h}'\\\sigma_{p}\phi\bar{h}'+2\,\phi^{2}\tan\left(\theta_{0}\right)-\sigma_{p}^{2}u'\end{pmatrix}
\end{equation}
and 
\begin{equation}
s_{q\,min}=\frac{\bar{h}_{q}}{2\left(1-\hat{q}^{2}\right)}.
\end{equation}


\begin{thebibliography}{52}
\expandafter\ifx\csname natexlab\endcsname\relax\def\natexlab#1{#1}\fi
\expandafter\ifx\csname bibnamefont\endcsname\relax
  \def\bibnamefont#1{#1}\fi
\expandafter\ifx\csname bibfnamefont\endcsname\relax
  \def\bibfnamefont#1{#1}\fi
\expandafter\ifx\csname citenamefont\endcsname\relax
  \def\citenamefont#1{#1}\fi
\expandafter\ifx\csname url\endcsname\relax
  \def\url#1{\texttt{#1}}\fi
\expandafter\ifx\csname urlprefix\endcsname\relax\def\urlprefix{URL }\fi
\providecommand{\bibinfo}[2]{#2}
\providecommand{\eprint}[2][]{\url{#2}}

\bibitem[{\citenamefont{Alberts et~al.}(2002)\citenamefont{Alberts, Johnson,
  Lewis, Raff, Roberts, and Walter}}]{Albertsbook2002}
\bibinfo{author}{\bibfnamefont{B.}~\bibnamefont{Alberts}},
  \bibinfo{author}{\bibfnamefont{A.}~\bibnamefont{Johnson}},
  \bibinfo{author}{\bibfnamefont{J.}~\bibnamefont{Lewis}},
  \bibinfo{author}{\bibfnamefont{M.}~\bibnamefont{Raff}},
  \bibinfo{author}{\bibfnamefont{K.}~\bibnamefont{Roberts}}, \bibnamefont{and}
  \bibinfo{author}{\bibfnamefont{P.}~\bibnamefont{Walter}},
  \emph{\bibinfo{title}{Molecular Biology of the Cell}}
  (\bibinfo{publisher}{Garland Science, New York}, \bibinfo{year}{2002}).

\bibitem[{\citenamefont{Seifert}(1997)}]{SeifertAiP1997}
\bibinfo{author}{\bibfnamefont{U.}~\bibnamefont{Seifert}},
  \bibinfo{journal}{Adv. Phys.} \textbf{\bibinfo{volume}{46}},
  \bibinfo{pages}{13} (\bibinfo{year}{1997}).

\bibitem[{\citenamefont{St\"{a}mpfli}(1958)}]{StampfliAABC1958}
\bibinfo{author}{\bibfnamefont{R.}~\bibnamefont{St\"{a}mpfli}},
  \bibinfo{journal}{An. Acad. Brasil. Cienc.} \textbf{\bibinfo{volume}{30}},
  \bibinfo{pages}{57} (\bibinfo{year}{1958}).

\bibitem[{\citenamefont{Neumann and Rosenheck}(1972)}]{NeumannJMB1972}
\bibinfo{author}{\bibfnamefont{E.}~\bibnamefont{Neumann}} \bibnamefont{and}
  \bibinfo{author}{\bibfnamefont{K.}~\bibnamefont{Rosenheck}},
  \bibinfo{journal}{J. Membr. Biol.} \textbf{\bibinfo{volume}{10}},
  \bibinfo{pages}{279} (\bibinfo{year}{1972}).

\bibitem[{\citenamefont{Abidor et~al.}(1979)\citenamefont{Abidor, Arakelyan,
  Chernomordik, Chizmadzhev, F., and Tarasevich}}]{Chizmadzhev1B&B1979}
\bibinfo{author}{\bibfnamefont{I.}~\bibnamefont{Abidor}},
  \bibinfo{author}{\bibfnamefont{V.~B.} \bibnamefont{Arakelyan}},
  \bibinfo{author}{\bibfnamefont{L.~V.} \bibnamefont{Chernomordik}},
  \bibinfo{author}{\bibfnamefont{Y.~A.} \bibnamefont{Chizmadzhev}},
  \bibinfo{author}{\bibfnamefont{P.~V.} \bibnamefont{F.}}, \bibnamefont{and}
  \bibinfo{author}{\bibfnamefont{M.~R.} \bibnamefont{Tarasevich}},
  \bibinfo{journal}{Bioelectrochem. Bioenerg.} \textbf{\bibinfo{volume}{6}},
  \bibinfo{pages}{37} (\bibinfo{year}{1979}).

\bibitem[{\citenamefont{Weaver}(2003)}]{WeaverIEEE2003}
\bibinfo{author}{\bibfnamefont{J.}~\bibnamefont{Weaver}},
  \bibinfo{journal}{IEEE Trans. Dielectr. Electr. Insul.}
  \textbf{\bibinfo{volume}{10}}, \bibinfo{pages}{754} (\bibinfo{year}{2003}).

\bibitem[{\citenamefont{Chen et~al.}(2006)\citenamefont{Chen, Smye, Robinson,
  and Evans}}]{ChenMBEC2006}
\bibinfo{author}{\bibfnamefont{C.}~\bibnamefont{Chen}},
  \bibinfo{author}{\bibfnamefont{S.}~\bibnamefont{Smye}},
  \bibinfo{author}{\bibfnamefont{M.}~\bibnamefont{Robinson}}, \bibnamefont{and}
  \bibinfo{author}{\bibfnamefont{J.}~\bibnamefont{Evans}},
  \bibinfo{journal}{Med. Biol. Eng. Comput.} \textbf{\bibinfo{volume}{44}},
  \bibinfo{pages}{5} (\bibinfo{year}{2006}).

\bibitem[{\citenamefont{Powell and Weaver}(1986)}]{WeaverB&B1986}
\bibinfo{author}{\bibfnamefont{K.~T.} \bibnamefont{Powell}} \bibnamefont{and}
  \bibinfo{author}{\bibfnamefont{J.~C.} \bibnamefont{Weaver}},
  \bibinfo{journal}{Bioelectrochem. Bioenerg.} \textbf{\bibinfo{volume}{15}},
  \bibinfo{pages}{211} (\bibinfo{year}{1986}).

\bibitem[{\citenamefont{Barnett and Weaver}(1991)}]{WeaverB&B1991}
\bibinfo{author}{\bibfnamefont{A.}~\bibnamefont{Barnett}} \bibnamefont{and}
  \bibinfo{author}{\bibfnamefont{J.~C.} \bibnamefont{Weaver}},
  \bibinfo{journal}{Bioelectrochem. Bioenerg.} \textbf{\bibinfo{volume}{25}},
  \bibinfo{pages}{163} (\bibinfo{year}{1991}).

\bibitem[{\citenamefont{DeBruin and Krassowska}(1999)}]{KrassowskaBioJ1999}
\bibinfo{author}{\bibfnamefont{K.~A.} \bibnamefont{DeBruin}} \bibnamefont{and}
  \bibinfo{author}{\bibfnamefont{W.}~\bibnamefont{Krassowska}},
  \bibinfo{journal}{Biophys. J.} \textbf{\bibinfo{volume}{77}},
  \bibinfo{pages}{1213} (\bibinfo{year}{1999}).

\bibitem[{\citenamefont{Joshi et~al.}(2001)\citenamefont{Joshi, Hu, Aly,
  Schoenbach, and Hjalmarson}}]{JoshiPRE2001}
\bibinfo{author}{\bibfnamefont{R.~P.} \bibnamefont{Joshi}},
  \bibinfo{author}{\bibfnamefont{Q.}~\bibnamefont{Hu}},
  \bibinfo{author}{\bibfnamefont{R.}~\bibnamefont{Aly}},
  \bibinfo{author}{\bibfnamefont{K.~H.} \bibnamefont{Schoenbach}},
  \bibnamefont{and} \bibinfo{author}{\bibfnamefont{H.~P.}
  \bibnamefont{Hjalmarson}}, \bibinfo{journal}{Phys. Rev. E}
  \textbf{\bibinfo{volume}{64}}, \bibinfo{pages}{011913}
  (\bibinfo{year}{2001}).

\bibitem[{\citenamefont{Joshi et~al.}(2002)\citenamefont{Joshi, Hu, Schoenbach,
  and Hjalmarson}}]{JoshiPRE2002}
\bibinfo{author}{\bibfnamefont{R.~P.} \bibnamefont{Joshi}},
  \bibinfo{author}{\bibfnamefont{Q.}~\bibnamefont{Hu}},
  \bibinfo{author}{\bibfnamefont{K.~H.} \bibnamefont{Schoenbach}},
  \bibnamefont{and} \bibinfo{author}{\bibfnamefont{H.~P.}
  \bibnamefont{Hjalmarson}}, \bibinfo{journal}{Phys. Rev. E}
  \textbf{\bibinfo{volume}{65}}, \bibinfo{pages}{041920}
  (\bibinfo{year}{2002}).

\bibitem[{\citenamefont{Neu and Krassowska}(2003)}]{KrassowskaPRE2003}
\bibinfo{author}{\bibfnamefont{J.~C.} \bibnamefont{Neu}} \bibnamefont{and}
  \bibinfo{author}{\bibfnamefont{W.}~\bibnamefont{Krassowska}},
  \bibinfo{journal}{Phys. Rev. E} \textbf{\bibinfo{volume}{67}},
  \bibinfo{pages}{021915} (\bibinfo{year}{2003}).

\bibitem[{\citenamefont{Smith et~al.}(2004)\citenamefont{Smith, Neu, and
  Krassowska}}]{KrassowskaBioJ2004}
\bibinfo{author}{\bibfnamefont{K.~C.} \bibnamefont{Smith}},
  \bibinfo{author}{\bibfnamefont{J.~C.} \bibnamefont{Neu}}, \bibnamefont{and}
  \bibinfo{author}{\bibfnamefont{W.}~\bibnamefont{Krassowska}},
  \bibinfo{journal}{Biophys. J.} \textbf{\bibinfo{volume}{86}},
  \bibinfo{pages}{2813} (\bibinfo{year}{2004}).

\bibitem[{\citenamefont{Krassowska and Filev}(2007)}]{KrassowskaBioJ2007}
\bibinfo{author}{\bibfnamefont{W.}~\bibnamefont{Krassowska}} \bibnamefont{and}
  \bibinfo{author}{\bibfnamefont{P.~D.} \bibnamefont{Filev}},
  \bibinfo{journal}{Biophys. J.} \textbf{\bibinfo{volume}{92}},
  \bibinfo{pages}{404} (\bibinfo{year}{2007}).

\bibitem[{\citenamefont{Bicout et~al.}(2006)\citenamefont{Bicout, Schmid, and
  Kats}}]{BicoutPRE2006}
\bibinfo{author}{\bibfnamefont{D.~J.} \bibnamefont{Bicout}},
  \bibinfo{author}{\bibfnamefont{F.}~\bibnamefont{Schmid}}, \bibnamefont{and}
  \bibinfo{author}{\bibfnamefont{E.}~\bibnamefont{Kats}},
  \bibinfo{journal}{Phys. Rev. E} \textbf{\bibinfo{volume}{73}},
  \bibinfo{pages}{060101(R)} (\bibinfo{year}{2006}).

\bibitem[{\citenamefont{Crowley}(1973)}]{CrowleyBioJ1973}
\bibinfo{author}{\bibfnamefont{J.~M.} \bibnamefont{Crowley}},
  \bibinfo{journal}{Biophys. J.} \textbf{\bibinfo{volume}{13}},
  \bibinfo{pages}{711} (\bibinfo{year}{1973}).

\bibitem[{\citenamefont{Lewis}(2003)}]{LewisIEEE2003}
\bibinfo{author}{\bibfnamefont{T.}~\bibnamefont{Lewis}}, \bibinfo{journal}{IEEE
  Trans. Dielectr. Electr. Insul.} \textbf{\bibinfo{volume}{10}},
  \bibinfo{pages}{769} (\bibinfo{year}{2003}).

\bibitem[{\citenamefont{Sens and Isambert}(2002)}]{IsambertPRL2002}
\bibinfo{author}{\bibfnamefont{P.}~\bibnamefont{Sens}} \bibnamefont{and}
  \bibinfo{author}{\bibfnamefont{H.}~\bibnamefont{Isambert}},
  \bibinfo{journal}{Phys. Rev. Lett.} \textbf{\bibinfo{volume}{88}},
  \bibinfo{pages}{128102} (\bibinfo{year}{2002}).

\bibitem[{\citenamefont{Movileanu et~al.}(2006)\citenamefont{Movileanu,
  Popescu, Ion, and Popescu}}]{PopescuBMB2006}
\bibinfo{author}{\bibfnamefont{L.}~\bibnamefont{Movileanu}},
  \bibinfo{author}{\bibfnamefont{D.}~\bibnamefont{Popescu}},
  \bibinfo{author}{\bibfnamefont{S.}~\bibnamefont{Ion}}, \bibnamefont{and}
  \bibinfo{author}{\bibfnamefont{A.}~\bibnamefont{Popescu}},
  \bibinfo{journal}{Bull. Math. Biol.} \textbf{\bibinfo{volume}{68}},
  \bibinfo{pages}{1231} (\bibinfo{year}{2006}).

\bibitem[{\citenamefont{Tieleman et~al.}(1997)\citenamefont{Tieleman, Marrink,
  and Berendsen}}]{TielemanBBA1997}
\bibinfo{author}{\bibfnamefont{D.~P.} \bibnamefont{Tieleman}},
  \bibinfo{author}{\bibfnamefont{S.~J.} \bibnamefont{Marrink}},
  \bibnamefont{and} \bibinfo{author}{\bibfnamefont{H.~J.~C.}
  \bibnamefont{Berendsen}}, \bibinfo{journal}{Biochim. Biophys. Acta, Rev. Biomembr.}
  \textbf{\bibinfo{volume}{1331}}, \bibinfo{pages}{235} (\bibinfo{year}{1997}).

\bibitem[{\citenamefont{Tarek}(2005)}]{TarekBioJ2005}
\bibinfo{author}{\bibfnamefont{M.}~\bibnamefont{Tarek}},
  \bibinfo{journal}{Biophys. J.} \textbf{\bibinfo{volume}{88}},
  \bibinfo{pages}{4045} (\bibinfo{year}{2005}).

\bibitem[{\citenamefont{Tieleman}(2004)}]{TielemanBMCBiochem2004}
\bibinfo{author}{\bibfnamefont{D.~P.} \bibnamefont{Tieleman}},
  \bibinfo{journal}{BMC Biochem.} \textbf{\bibinfo{volume}{5}},
  \bibinfo{pages}{10} (\bibinfo{year}{2004}).

\bibitem[{\citenamefont{Gurtovenko and Vattulainen}(2007)}]{GurtovenkoBioJ2007}
\bibinfo{author}{\bibfnamefont{A.~A.} \bibnamefont{Gurtovenko}}
  \bibnamefont{and}
  \bibinfo{author}{\bibfnamefont{I.}~\bibnamefont{Vattulainen}},
  \bibinfo{journal}{Biophys. J.} \textbf{\bibinfo{volume}{92}},
  \bibinfo{pages}{1878} (\bibinfo{year}{2007}).

\bibitem[{\citenamefont{Bier et~al.}(2002)\citenamefont{Bier, Chen,
  Gowrishankar, Astumian, and Lee}}]{BierPRE2002}
\bibinfo{author}{\bibfnamefont{M.}~\bibnamefont{Bier}},
  \bibinfo{author}{\bibfnamefont{W.}~\bibnamefont{Chen}},
  \bibinfo{author}{\bibfnamefont{T.~R.} \bibnamefont{Gowrishankar}},
  \bibinfo{author}{\bibfnamefont{R.~D.} \bibnamefont{Astumian}},
  \bibnamefont{and} \bibinfo{author}{\bibfnamefont{R.~C.} \bibnamefont{Lee}},
  \bibinfo{journal}{Phys. Rev. E} \textbf{\bibinfo{volume}{66}},
  \bibinfo{pages}{062905} (\bibinfo{year}{2002}).

\bibitem[{\citenamefont{Melikov et~al.}(2001)\citenamefont{Melikov, Frolov,
  Shcherbakov, Samsonov, Chizmadzhev, and Chernomordik}}]{ChizmadzhevBioJ2001}
\bibinfo{author}{\bibfnamefont{K.~C.} \bibnamefont{Melikov}},
  \bibinfo{author}{\bibfnamefont{V.~A.} \bibnamefont{Frolov}},
  \bibinfo{author}{\bibfnamefont{A.}~\bibnamefont{Shcherbakov}},
  \bibinfo{author}{\bibfnamefont{A.~V.} \bibnamefont{Samsonov}},
  \bibinfo{author}{\bibfnamefont{Y.~A.} \bibnamefont{Chizmadzhev}},
  \bibnamefont{and} \bibinfo{author}{\bibfnamefont{L.~V.}
  \bibnamefont{Chernomordik}}, \bibinfo{journal}{Biophys. J.}
  \textbf{\bibinfo{volume}{80}}, \bibinfo{pages}{1829} (\bibinfo{year}{2001}).

\bibitem[{\citenamefont{Kakorin et~al.}(1998)\citenamefont{Kakorin, Redeker,
  and Neumann}}]{NeumannEBJ1998}
\bibinfo{author}{\bibfnamefont{S.}~\bibnamefont{Kakorin}},
  \bibinfo{author}{\bibfnamefont{E.}~\bibnamefont{Redeker}}, \bibnamefont{and}
  \bibinfo{author}{\bibfnamefont{E.}~\bibnamefont{Neumann}},
  \bibinfo{journal}{Eur. Biophys. J.} \textbf{\bibinfo{volume}{27}},
  \bibinfo{pages}{43} (\bibinfo{year}{1998}).

\bibitem[{\citenamefont{Tekle et~al.}(2001)\citenamefont{Tekle, Astumian,
  Friauf, and Chock}}]{TekleBioJ2001}
\bibinfo{author}{\bibfnamefont{E.}~\bibnamefont{Tekle}},
  \bibinfo{author}{\bibfnamefont{R.~D.} \bibnamefont{Astumian}},
  \bibinfo{author}{\bibfnamefont{W.~A.} \bibnamefont{Friauf}},
  \bibnamefont{and} \bibinfo{author}{\bibfnamefont{P.~B.} \bibnamefont{Chock}},
  \bibinfo{journal}{Biophys. J.} \textbf{\bibinfo{volume}{81}},
  \bibinfo{pages}{960} (\bibinfo{year}{2001}).

\bibitem[{\citenamefont{Riske and Dimova}(2005)}]{RiskeBioJ2005}
\bibinfo{author}{\bibfnamefont{K.~A.} \bibnamefont{Riske}} \bibnamefont{and}
  \bibinfo{author}{\bibfnamefont{R.}~\bibnamefont{Dimova}},
  \bibinfo{journal}{Biophys. J.} \textbf{\bibinfo{volume}{88}},
  \bibinfo{pages}{1143} (\bibinfo{year}{2005}).

\bibitem[{\citenamefont{Dimova et~al.}(2007)\citenamefont{Dimova, Riske,
  Aranda, Bezlyepkina, Knorr, and Lipowsky}}]{DimovaSM2007}
\bibinfo{author}{\bibfnamefont{R.}~\bibnamefont{Dimova}},
  \bibinfo{author}{\bibfnamefont{K.}~\bibnamefont{Riske}},
  \bibinfo{author}{\bibfnamefont{S.}~\bibnamefont{Aranda}},
  \bibinfo{author}{\bibfnamefont{N.}~\bibnamefont{Bezlyepkina}},
  \bibinfo{author}{\bibfnamefont{R.}~\bibnamefont{Knorr}}, \bibnamefont{and}
  \bibinfo{author}{\bibfnamefont{R.}~\bibnamefont{Lipowsky}},
  \bibinfo{journal}{Soft matter} \textbf{\bibinfo{volume}{3}},
  \bibinfo{pages}{817} (\bibinfo{year}{2007}).

\bibitem[{\citenamefont{Huang}(1986)}]{HuangBioJ1986}
\bibinfo{author}{\bibfnamefont{H.~W.} \bibnamefont{Huang}},
  \bibinfo{journal}{Biophys. J.} \textbf{\bibinfo{volume}{50}},
  \bibinfo{pages}{1061} (\bibinfo{year}{1986}).

\bibitem[{\citenamefont{Seifert and Langer}(1993)}]{SeifertEPL1993}
\bibinfo{author}{\bibfnamefont{U.}~\bibnamefont{Seifert}} \bibnamefont{and}
  \bibinfo{author}{\bibfnamefont{S.~A.} \bibnamefont{Langer}},
  \bibinfo{journal}{Europhys. Lett.} \textbf{\bibinfo{volume}{23}},
  \bibinfo{pages}{71} (\bibinfo{year}{1993}).

\bibitem[{\citenamefont{Goldstein et~al.}(1996)\citenamefont{Goldstein, Nelson,
  Powers, and Seifert}}]{GoldsteinJPF1996}
\bibinfo{author}{\bibfnamefont{R.}~\bibnamefont{Goldstein}},
  \bibinfo{author}{\bibfnamefont{P.}~\bibnamefont{Nelson}},
  \bibinfo{author}{\bibfnamefont{T.}~\bibnamefont{Powers}}, \bibnamefont{and}
  \bibinfo{author}{\bibfnamefont{U.}~\bibnamefont{Seifert}},
  \bibinfo{journal}{J. Phys. II France} \textbf{\bibinfo{volume}{6}},
  \bibinfo{pages}{767} (\bibinfo{year}{1996}).

\bibitem[{\citenamefont{Evans and Simon}(1975)}]{EvansBioJ1975}
\bibinfo{author}{\bibfnamefont{E.}~\bibnamefont{Evans}} \bibnamefont{and}
  \bibinfo{author}{\bibfnamefont{S.}~\bibnamefont{Simon}},
  \bibinfo{journal}{Biophys. J.} \textbf{\bibinfo{volume}{15}},
  \bibinfo{pages}{850} (\bibinfo{year}{1975}).

\bibitem[{\citenamefont{Requena et~al.}(1975)\citenamefont{Requena, Haydon, and
  Hladky}}]{RequenaBioJ1975}
\bibinfo{author}{\bibfnamefont{J.}~\bibnamefont{Requena}},
  \bibinfo{author}{\bibfnamefont{D.}~\bibnamefont{Haydon}}, \bibnamefont{and}
  \bibinfo{author}{\bibfnamefont{S.}~\bibnamefont{Hladky}},
  \bibinfo{journal}{Biophys. J.} \textbf{\bibinfo{volume}{15}},
  \bibinfo{pages}{77} (\bibinfo{year}{1975}).

\bibitem[{\citenamefont{Andrews et~al.}(1970)\citenamefont{Andrews, Manev, and
  Haydon}}]{AndrewsSDFS1970}
\bibinfo{author}{\bibfnamefont{D.}~\bibnamefont{Andrews}},
  \bibinfo{author}{\bibfnamefont{E.}~\bibnamefont{Manev}}, \bibnamefont{and}
  \bibinfo{author}{\bibfnamefont{D.}~\bibnamefont{Haydon}},
  \bibinfo{journal}{Spec. Discuss. Faraday Soc.} \textbf{\bibinfo{volume}{1}},
  \bibinfo{pages}{46} (\bibinfo{year}{1970}).

\bibitem[{\citenamefont{Raudino and Mauzerall}(1986)}]{RaudinoBioJ1986}
\bibinfo{author}{\bibfnamefont{A.}~\bibnamefont{Raudino}} \bibnamefont{and}
  \bibinfo{author}{\bibfnamefont{D.}~\bibnamefont{Mauzerall}},
  \bibinfo{journal}{Biophys. J} \textbf{\bibinfo{volume}{50}},
  \bibinfo{pages}{441} (\bibinfo{year}{1986}).

\bibitem[{\citenamefont{B\"{o}ckmann et~al.}(2008)\citenamefont{B\"{o}ckmann,
  de~Groot, Kakorin, Neumann, and Grubm\"{u}ller}}]{BockmannBioJ2008}
\bibinfo{author}{\bibfnamefont{R.~A.} \bibnamefont{B\"{o}ckmann}},
  \bibinfo{author}{\bibfnamefont{B.~L.} \bibnamefont{de~Groot}},
  \bibinfo{author}{\bibfnamefont{S.}~\bibnamefont{Kakorin}},
  \bibinfo{author}{\bibfnamefont{E.}~\bibnamefont{Neumann}}, \bibnamefont{and}
  \bibinfo{author}{\bibfnamefont{H.}~\bibnamefont{Grubm\"{u}ller}},
  \bibinfo{journal}{Biophys. J.} \textbf{\bibinfo{volume}{95}},
  \bibinfo{pages}{1837} (\bibinfo{year}{2008}).

\bibitem[{\citenamefont{Chen et~al.}(1995)\citenamefont{Chen, Lubensky, and
  MacKintosh}}]{ChenPRE1995}
\bibinfo{author}{\bibfnamefont{C.-M.} \bibnamefont{Chen}},
  \bibinfo{author}{\bibfnamefont{T.~C.} \bibnamefont{Lubensky}},
  \bibnamefont{and} \bibinfo{author}{\bibfnamefont{F.~C.}
  \bibnamefont{MacKintosh}}, \bibinfo{journal}{Phys. Rev. E}
  \textbf{\bibinfo{volume}{51}}, \bibinfo{pages}{504} (\bibinfo{year}{1995}).

\bibitem[{\citenamefont{Chen and MacKintosh}(1996)}]{ChenPRE1996}
\bibinfo{author}{\bibfnamefont{C.-M.} \bibnamefont{Chen}} \bibnamefont{and}
  \bibinfo{author}{\bibfnamefont{F.~C.} \bibnamefont{MacKintosh}},
  \bibinfo{journal}{Phys. Rev. E} \textbf{\bibinfo{volume}{53}},
  \bibinfo{pages}{4933} (\bibinfo{year}{1996}).

\bibitem[{\citenamefont{Lubensky and MacKintosh}(1993)}]{LubenskyPRL1993}
\bibinfo{author}{\bibfnamefont{T.~C.} \bibnamefont{Lubensky}} \bibnamefont{and}
  \bibinfo{author}{\bibfnamefont{F.~C.} \bibnamefont{MacKintosh}},
  \bibinfo{journal}{Phys. Rev. Lett.} \textbf{\bibinfo{volume}{71}},
  \bibinfo{pages}{1565} (\bibinfo{year}{1993}).

\bibitem[{\citenamefont{Andelman et~al.}(1987)\citenamefont{Andelman, Brochard,
  and Joanny}}]{AndelmanJCP1986}
\bibinfo{author}{\bibfnamefont{D.}~\bibnamefont{Andelman}},
  \bibinfo{author}{\bibfnamefont{F.}~\bibnamefont{Brochard}}, \bibnamefont{and}
  \bibinfo{author}{\bibfnamefont{J.-F.} \bibnamefont{Joanny}},
  \bibinfo{journal}{J. Chem. Phys.} \textbf{\bibinfo{volume}{86}},
  \bibinfo{pages}{3673} (\bibinfo{year}{1987}).

\bibitem[{\citenamefont{Brochard and Lennon}(1975)}]{BrochardJDP1975}
\bibinfo{author}{\bibfnamefont{F.}~\bibnamefont{Brochard}} \bibnamefont{and}
  \bibinfo{author}{\bibfnamefont{J.}~\bibnamefont{Lennon}},
  \bibinfo{journal}{J. Phys. (Paris)} \textbf{\bibinfo{volume}{36}},
  \bibinfo{pages}{1035} (\bibinfo{year}{1975}).

\bibitem[{\citenamefont{Lacoste et~al.}(2007)\citenamefont{Lacoste,
  Lagomarsino, and Joanny}}]{LacosteEPL2007}
\bibinfo{author}{\bibfnamefont{D.}~\bibnamefont{Lacoste}},
  \bibinfo{author}{\bibfnamefont{M.~C.} \bibnamefont{Lagomarsino}},
  \bibnamefont{and} \bibinfo{author}{\bibfnamefont{J.~F.}
  \bibnamefont{Joanny}}, \bibinfo{journal}{Europhys. Lett.}
  \textbf{\bibinfo{volume}{77}}, \bibinfo{pages}{18006} (\bibinfo{year}{2007}).

\bibitem[{\citenamefont{Ajdari}(1995)}]{AjdariPRL1995}
\bibinfo{author}{\bibfnamefont{A.}~\bibnamefont{Ajdari}},
  \bibinfo{journal}{Phys. Rev. Lett.} \textbf{\bibinfo{volume}{75}},
  \bibinfo{pages}{755} (\bibinfo{year}{1995}).

\bibitem[{\citenamefont{Pandit et~al.}(2003)\citenamefont{Pandit, Bostick, and
  Berkowitz}}]{BerkowitzBioJ2003}
\bibinfo{author}{\bibfnamefont{S.~A.} \bibnamefont{Pandit}},
  \bibinfo{author}{\bibfnamefont{D.}~\bibnamefont{Bostick}}, \bibnamefont{and}
  \bibinfo{author}{\bibfnamefont{M.~L.} \bibnamefont{Berkowitz}},
  \bibinfo{journal}{Biophys. J.} \textbf{\bibinfo{volume}{84}},
  \bibinfo{pages}{3743} (\bibinfo{year}{2003}).

\bibitem[{\citenamefont{Saiz and Klein}(2002)}]{KleinJCP2002}
\bibinfo{author}{\bibfnamefont{L.}~\bibnamefont{Saiz}} \bibnamefont{and}
  \bibinfo{author}{\bibfnamefont{M.~L.} \bibnamefont{Klein}},
  \bibinfo{journal}{J. Chem. Phys.} \textbf{\bibinfo{volume}{116}},
  \bibinfo{pages}{3052} (\bibinfo{year}{2002}).

\bibitem[{\citenamefont{Nelson}(2004)}]{Nelsonbook2004}
\bibinfo{author}{\bibfnamefont{P.}~\bibnamefont{Nelson}},
  \emph{\bibinfo{title}{Biological Physics}} (\bibinfo{publisher}{Freeman, New
  York}, \bibinfo{year}{2004}).

\bibitem[{\citenamefont{Kotulska et~al.}(2007)\citenamefont{Kotulska, Kubica,
  Koronkiewicz, and Kalinowski}}]{KotulskaBioelectro2007}
\bibinfo{author}{\bibfnamefont{M.}~\bibnamefont{Kotulska}},
  \bibinfo{author}{\bibfnamefont{K.}~\bibnamefont{Kubica}},
  \bibinfo{author}{\bibfnamefont{S.}~\bibnamefont{Koronkiewicz}},
  \bibnamefont{and}
  \bibinfo{author}{\bibfnamefont{S.}~\bibnamefont{Kalinowski}},
  \bibinfo{journal}{Bioelectrochemistry} \textbf{\bibinfo{volume}{70}},
  \bibinfo{pages}{64 } (\bibinfo{year}{2007}).

\bibitem[{\citenamefont{Evans et~al.}(1976)\citenamefont{Evans, Waugh, and
  Melnik}}]{EvansBioJ1976}
\bibinfo{author}{\bibfnamefont{E.~A.} \bibnamefont{Evans}},
  \bibinfo{author}{\bibfnamefont{R.}~\bibnamefont{Waugh}}, \bibnamefont{and}
  \bibinfo{author}{\bibfnamefont{L.}~\bibnamefont{Melnik}},
  \bibinfo{journal}{Biophys. J.} \textbf{\bibinfo{volume}{16}},
  \bibinfo{pages}{585} (\bibinfo{year}{1976}).

\bibitem[{\citenamefont{Freeman et~al.}(1994)\citenamefont{Freeman, Wang, and
  Weaver}}]{FreemanBioJ1994}
\bibinfo{author}{\bibfnamefont{S.}~\bibnamefont{Freeman}},
  \bibinfo{author}{\bibfnamefont{M.}~\bibnamefont{Wang}}, \bibnamefont{and}
  \bibinfo{author}{\bibfnamefont{J.}~\bibnamefont{Weaver}},
  \bibinfo{journal}{Biophys. J.} \textbf{\bibinfo{volume}{67}},
  \bibinfo{pages}{42 } (\bibinfo{year}{1994}).

\bibitem[{\citenamefont{Hohenberg and Halperin}(1977)}]{HohenbergRoMP1977}
\bibinfo{author}{\bibfnamefont{P.~C.} \bibnamefont{Hohenberg}}
  \bibnamefont{and} \bibinfo{author}{\bibfnamefont{B.~I.}
  \bibnamefont{Halperin}}, \bibinfo{journal}{Rev. Mod. Phys.}
  \textbf{\bibinfo{volume}{49}}, \bibinfo{pages}{435} (\bibinfo{year}{1977}).

\end{thebibliography}
\end{document}